\documentclass{article}
\usepackage[utf8]{inputenc}
\usepackage[margin=1.4in]{geometry}

\usepackage{graphicx}
\usepackage{amsmath, amsfonts}
\usepackage{setspace}
\usepackage{multirow}
\usepackage{graphicx, epstopdf}
\usepackage{mathrsfs}
\usepackage{mdwlist}
\usepackage{caption}
\usepackage{rotating}
\usepackage{footnote}
\usepackage{lscape}
\usepackage[table,xcdraw]{xcolor}
\newcommand{\test}[1]{\expandafter\hat#1} 
\usepackage{algorithm}
\usepackage{algpseudocode}
\usepackage{bm}
\usepackage{rotating}
\usepackage{pdflscape}
\usepackage{subfig}
\usepackage[numbers]{natbib}
\usepackage[countmax]{subfloat}

\title{Surrogate-based Optimization using Mutual Information for Computer Experiments (optim-MICE)}
\author{Theodoros Mathikolonis \footnote{Department of Statistical Science, University College London, UK (theodoros.mathikolonis.13@ucl.ac.uk)} \,\,\,\ Serge Guillas \footnote{Department of Statistical Science, University College London, UK (s.guillas@ucl.ac.uk)}}

\begin{document}

\maketitle

\begin{abstract}
The computational burden of running a complex computer model can make optimization impractical. Gaussian Processes (GPs) are statistical surrogates (also known as emulators) that alleviate this issue since they cheaply replace the computer model. As a result, the exploration vs. exploitation trade-off strategy can be accelerated by building a GP surrogate. In this paper, we propose a new surrogate-based optimization scheme that minimizes the number of evaluations of the computationally expensive function. Taking advantage of parallelism of the evaluation of the unknown function, the uncertain regions are explored simultaneously, and a batch of input points is chosen using Mutual Information for Computer Experiments (MICE), a sequential design algorithm which maximize the information theoretic Mutual Information over the input space. The computational efficiency of interweaving the optimization scheme with MICE (optim-MICE) is examined and demonstrated on test functions. Optim-MICE is compared with state-of-the-art heuristics such as Efficient Global
Optimization (EGO) and GP-Upper Confidence Bound (GP-UCB). We demonstrate that optim-MICE outperforms these schemes on a large range of computational experiments.
\end{abstract}

\section{Introduction}
\label{intro}
Computer models, also known as simulators, are widely used to study physical processes. When run at high fidelity, simulators can become computationally expensive. Optimizing an unknown function $f$ from a set of sequential evaluations is a common task in different fields of science and engineering. The 'standard' mathematical approach in an optimization is to use all the available information contained in the derivatives of any function. However, this is not always the case as many practical applications require to optimize a function $f$ over a domain of interest where derivatives are unavailable, unreliable or computationally prohibitive. For instance, $f$ can be very expensive to compute or may have discontinuous derivatives. These problems are usually referred as a derivative-free optimization or black-box optimization since the analytic form of the function is not known.\\
\\
The development of different derivative-free algorithms starts in the mid-1960s when the Nelder-Mead (NM) simplex algorithm is invented. The NM algorithm is the most widely used direct search method for solving unconstrained optimization without derivatives \citep{NM1965}. Since then, a lot of studies have been done in the area with significant improvements \citep{boukouvala2016global,Kolda2003,Rios2013}. Many derivative-free algorithms, such as the Genetic Algorithm, Random Search method, and Particle Swarm optimization have been proved to be reliable techniques for finding the global optimum. However, they often need a large number of function evaluations, and therefore a lot of computational resources, which renders them unaffordable for computationally expensive problems.\\
\\
Nowadays, surrogate-based optimization strategies are intensively used due to their efficiency in solving a computational expensive optimization problem by reducing the number of function evaluations. During the optimization procedure, a statistical surrogate model replaces and accurately represents, the computer model. It is mainly used for gaining insights about the characteristics of the unknown function and making fast predictions without evaluating the computationally expensive simulator. A variety of surrogate-based optimization techniques have been proposed in the literature \citep{cai2017multi,forrester2009recent,gutmann2001radial,huang2006,jakobsson2010method,Lee2019,marsden2004optimal,regis2007improved}. Kriging, also known as Gaussian Process (GP) modelling, is the most popular and commonly used statistical surrogate model due to its ability to effectively provide an uncertainty estimation of the prediction. Kriging-based optimization techniques have been used to deal with constrained problems \citep{sasena2002exploration,li2017kriging,parr2012infill,wang2018constrained}, single objective problems \citep{dong2019multi,huang2006sequential,huang2006,jones1998} and multi-objective problems \citep{couckuyt2014fast,feliot2017bayesian,knowles2006parego}.\\
\\
The most successful kriging-based optimization technique is the Efficient Global Optimization (EGO) algorithm proposed in \citep{jones1998}. At the beginning, a GP model is fitted based on an initial design set. Then, the algorithm follows an iterative procedure where, at each iteration, a new candidate point is chosen by maximizing a sampling criterion called Expected Improvement (EI) and then, the GP is re-fitted again considering the updated design set. The traditional EGO algorithm is also extended into a parallel optimization scheme, by which multiple candidate points are chosen at each iteration. Such a strategy is the \textit{q}-EI, proposed in \citep{ginsbourger2009}, where a batch of \textit{q} points are added to the design set, at the same time, by maximizing an approximate expression of EI. Different developments of the parallel EGO algorithm can be also found in \citep{feng2015multiobjective,sobester2004parallel,zhan2017pseudo}. \\
\\
Srinivas et al. \citep{srinivas2010} developed a new GP optimization algorithm that chooses the new candidate points that maximize the upper confidence bound (UCB). The optimization of the unknown function $f$ is formalized as a multi-armed bandit problem where the GP predictive uncertainty is used to control the exploration and exploitation. The performance of the GP-UCB algorithm is measured according to regret, the difference between the actual maximum and the best result achieved, or the cumulative regret (sum of the regrets), the loss incurred due to not knowing the $f$'s maximum value. The objective is to minimize the cumulative regret or maximize the sum of rewards, which essentially is the same as maximizing the black-box function. In contrast to EGO-based algorithms where the convergence rates remain elusive \citep{bull2011convergence,vazquez2010convergence}, Srinivas et al. \citep{srinivas2010} give the first theoretical bounds of cumulative regret, for functions sampled from a GP, which can be translated into convergence rates for GP optimization. A technical connection between the multi-armed bandit setting and the experimental design is also achieved as the regret is bounded by an information gain quantity used as a sampling criterion \citep{krause2012near}. A theoretical analysis under the multi-armed bandit setting can be found in previous studies but these were not applicable in a GP framework \citep{auer2002,dani2007,kleinberg2008,lai1985}. \\
\\
The trade-off between exploration and exploitation for the contextual GP bandit problems is addressed in the  Contextual Gaussian Process Bandit Optimization (CGP-UCB) algorithm proposed by Krause et al. \citep{krause2011contextual}. The pay-off function corresponding to context-action pairs is modelled as a sample from a GP over the context-action space. A multi-fidelity version of the GP bandit problem was investigated by Kandasamy et al. \citep{kandasamy2016gaussian} where the MF-GP-UCB algorithm, an extension of GP-UCB, aims to eliminate the low function value regions using cheap lower fidelities and focus on a small, but promising, region using a sequence of successively higher fidelities. When the unknown function is sufficiently smooth and the set $\mathcal{X}$ compact and convex, $\mathcal{X} \subset \mathbb{R}^d$, the performance of GP-UCB is reduced due to the increase in cardinality $|\mathcal{X}|$. To overcome this issue and control the discretization error, it has been proved
that GP-UCB can be run using the settings for finite set \citep{srinivas2012}. An improved version of the GP-UCB is also suggested by Contal and Vayatis \citep{contal2016} where, to precisely control the discretization error, a sequence of uniform discretizations are constructed using genering chaining that leads to tight bounds. \\
\\
Following Srinivas et al. \citep{srinivas2010} approach, and under the exploration-exploitation framework, different sequential optimization schemes have been proposed in the literature, which incorporate a parallel strategy: multiple evaluations are performed in parallel whereby a batch of multiple input points is selected at each iteration. A batch optimization strategy was first introduced by Azimi et al. \citep{azimi2010}, using the Monte-Carlo as an alternative to GP, based on the idea of simulation matching. Choosing the level of parallelism and whether to sequentially evaluate the function or evaluating it on a batch mode, was studied by Azimi et al. \citep{azimi2012}: the batch size is adaptively changing based on the expected prediction error. The connection between the multi-armed bandit and the experimental design was introduced by Srinivas et al. \citep{srinivas2010} and extended by Desautels et al. \citep{desautels2014}. Precisely, two algorithms accommodate the parallel strategy and the batch execution: the GP-BUCB, which selects at each iteration a batch of fixed size, and GP-AUCB, a variant of the first algorithm, which adaptively exploits parallelism to choose a batch of input points where its size is based on the amount of information gained about the unknown function. The cumulative regret bounds are provided for both algorithms. \\
\\
By combining two strategies to determine the input points for each batch of a fixed size, the GP-UCB-PE algorithm, proposed by Contal et al. \citep{contal2013}, aims to maximize an unknown function with the lowest possible number of function evaluations. Specifically, the UCB policy is used to select the first input point by balancing the exploration and exploitation whereas the remaining input points are chosen using the Pure Exploration (PE) strategy from regions which contain the true optimum with high probability. The PE step follows a greedy strategy where the input points are chosen one by one based on the Advanced Learning MacKay (ALM) \citep{mackay1992}, an adaptive experimental design which maximizes the information gained about the unknown function. As a result, only the input points that maximize the information gain quantity, as in the work of Srinivas et al \citep{srinivas2010} and Desautels et al. \citep{desautels2014}, are chosen. Under the GP framework, the information gain quantity is computed based on the predictive variance.\\
\\
In this paper we propose a new surrogate-based optimization scheme which is built upon Contal et al. \citep{contal2013} and Beck and Guillas \citep{Joakim2016}. It aims to maximize a complex and a computational expensive function with the lowest possible number of function evaluations. As in Contal et al. \citep{contal2013}, a batch of input points is chosen, at each iteration, in a two-step process. The first input point is chosen based on the UCB policy whereas the remaining points are selected via the PE strategy, one by one. The PE is based on MICE (Mutual Information for Computer Experiments), an adaptive experimental design \citep{Joakim2016} which is the improved version of the sequential Mutual Information (MI) algorithm \citep{krause2008}. In the PE step, the uncertain regions are explored and the input points chosen are the ones that maximize an information theoretic mutual information measure. \\
\\
The paper is organized as follow. Section 2 states the problem and gives the background as well as some basic knowledge about surrogate models and experimental design. A description of the optim-MICE algorithm is given in Section 3. The optim-MICE algorithm is compared with other related optimization methods on different computational experiments. The algorithm settings used in each technique and computational experiment are shown in Section 4. The results and discussion about the computational efficiency, and performance of the optim-MICE algorithm compared with the alternative methods, are presented in Section 5. We finally conclude in Section 6. 

\section{Problem Statement and Background}
\label{sec:1}
The current study addresses the problem of sequentially optimizing an unknown function. Let $f :\mathcal{X} \to \mathbb{R}$ be our unknown function with $\mathcal{X} \subseteq \mathbb{R}^d$, compact and convex. The aim is to find with the lowest possible number of function evaluations, the maximum of the unknown function

\begin{equation}
f(x^*) = \underset{x \,\in \,\mathcal{X}}{\max} \,\,  f(x),
\end{equation}\
\\
where $x^*$ denotes the true location of the maximum of $f$. At each iteration $t$, a batch of $K$ input points ($x_t^k$) in $\mathcal{X}$ are chosen and then, the function values at these locations are simultaneously obtained. \\
\\
A common sequential design strategy to optimise a black-box function is the Bayesian Optimization where a probabilistic model is firstly built and then, an acquisition function is used to determine the new input points by satisfying some optimality criterion. Although the Bayesian philosophy is not adopted in this study, the Bayesian Optimization framework is followed to tackle the sequential batch optimization problem. A well known probabilistic model used in black-box optimization is the GP regression due to its flexibility and tractability \citep{snoek2012}. \\
\\
A GP regression is also widely used as a statistical surrogate model, also known as emulator or meta-model. Indeed computer models are often complex and  computational expensive black-box functions and thus make activities such as optimization, sensitivity and uncertainty analyses impractical. To overcome this issue, a statistical surrogate model is used as a mean for designing and analysing computer experiments. Surrogates models, and more precisely GP emulators, can represent complicated functional forms with the aim to approximate the input-output behaviour and accurately represent the analytical model even with extensions to time series of outputs \citep{ohagan2006,sarri2012,guillas2018functional}.

\subsection{Gaussian Process}
A GP enforces implicit properties of the unknown function $f$ without relying on any parametric assumptions. By modelling $f$ as a sample from a GP, a certain level of smoothness, and correlation between nearby locations can be formalized. A GP is a continuous extension of multidimensional normal distribution, defined by the mean function $m(x)$ and the covariance function $c(x,x')$. The unknown function $f$ can be thought as a Gaussian random function and its output $y=f(x)$ can be described as GP:
\begin{align}
f(x) &\sim GP(m(x), c(x,x')),\\
\text{where}\,\,\,\,\ m(x) &= \mathbb{E}[f(x)],\\
\text{and}\,\,\,\,\ c(x,x') &= \mathbb{E}\left[\left(f(x) - m(x)\right)\left(f(x') - m(x')\right)\right].
\end{align}
The covariance function is defined as $c(x,x') = \sigma^2K(x,x')$: a product of the process variance ($\sigma^2 > 0$) and a correlation matrix $K$. The mean function is often set to zero so as to let the variation in the data be explained by the covariance function \citep{rasmussen2005}. The covariance function - also known as kernel - is at the heart of the GP as it encodes the properties of $f$. Unlike the mean function which can be chosen freely, an arbitrary covariance function, in general, is not valid as it has to be a symmetric positive semi-definite function. A range of the different class of covariance functions can be found in Rasmussen et al. \citep{rasmussen2005} but the most commonly-used are: 
\begin{itemize}
\item \textit{Separable Power Exponential} with $\xi = (l_1, \dots, l_d)^T \in \mathbb{R}^{d}_{+}$ where the length-scale for the \textit{i}th input dimension is $l_i >0$. The degree of smoothness is controlled by $0 < p \leq 2$ with a typical default choice 2 \citep{santner2003,gramacy2012}, 
\begin{equation}
K(x,x'|\xi)) =  \prod_{i=1}^{d} \exp \left\{ - \frac{\|x-x'\|^{p}}{l_i} \right\},
\end{equation}

\item \textit{Mat\'ern} with length-scale $l_i>0$, $\xi = (l_1, \dots, l_d)^T$, and parameter $\nu$ which controls the smoothness level,
\begin{equation}
K^*(x,x'|\xi,\nu) = \prod_{i=1}^{d}\frac{1}{2^{\nu-1}\Gamma(\nu)} \left(\frac{2\nu^{1/2} \| x - x'\|}{l_i}\right)^\nu J_\nu \left(\frac{2\nu^{1/2} \|x - x'\|}{l_i} \right),  
\end{equation} 
where $\Gamma_\nu$ is the Gamma function for $\nu$ and $J_{\nu}$ is a modified Bessel function of order $\nu > 0$. A common choice for the degree of smoothness are $\nu = 3/2$ and $\nu = 5/2$ \citep{stein1999,minasny2005,rougier2009}.
\end{itemize} \
\\
The uncertain parameters can be modelled using a Bayesian approach \citep{gramacy2009,handcock1993,rasmussen2005}. However, the current study adopts the Design and Analysis of Computer Experiments (DACE) framework proposed by Sacks et al. \citep{sacks1989}. The GP regression not only offers a prediction at a new input point but also provides an estimate of the uncertainty in that prediction \citep{sacks1989}. Conditionally on the training outputs after $T$ iterations,  $Y_T = \left[y_t, \dots, y_T \right]^T$ at points $X_T=\left\{x_t, \dots, x_T \right\}$, the process is still a GP and the predictive distribution of the output at a new input points $x$, also known as test points, is a multivariate normal with mean $\test{y}(x) $ and variance $\test{s}^2(x)$: 
\begin{align} 
\test{y}_{T+1}(x) &= k_T(x)^T(K_T + \sigma^2I)^{-1}Y_T \label{eq1} \\
\text{and} \,\ \test{s}_{T+1}^2(x) &= k(x, x)-k_T(x)^T(K_T+\sigma^2I)^{-1}k_T(x) \label{eq2}
\end{align}  
where $k_T(x) = \left[k(x_1, x), \dots, k(x_T,x)\right]^T$ is the vector of covariances between the input points already chosen and $x$ and $K_T=\left[(x,x')\right]_{x,x' \in X_T}$ is the covariance matrix. 

\subsection{Experimental Design}
The computational complexity of a computer model often only allows to perform a limited number of runs. To minimize the computational cost, and maximize the information gained about the unknown function, an experimental design is used where an efficient set of input points ($n$ design points) is chosen strategically based on various optimization techniques \citep{santner2003,sacks1989,johnson1990}. The experimental design strategies are classified mainly into two categories: space-filling designs and adaptive designs.\\
\\
Space-filling designs choose all the design points before computing any function evaluation. All regions of the design space are treated as equally important and, as a result, a certain amount of computational time is wasted because unnecessary  regions are explored. Examples of space-filling designs are uniform designs, maximin and minimax distance design and Latin Hypercube designs (LHD) \citep{sacks1989,pronzato2012,simpson2001}. The current study focuses on the adaptive designs which, compared with space-filling designs, can be sometimes computationally expensive but often more effective \citep{Joakim2016}. In adaptive designs, only the most informative input points are included in the training data set by optimizing, at each step of the experimental design process, a specific design criterion \citep{Joakim2016,santner2003,gramacy2009,chen2008}. The design points are chosen sequentially, often one-at-a time or in batches, from regions where uncertainty is large. The adaptive designs described below measure an information gain quantity which can serve as a criterion for designing computer experiments \citep{currin1988}. For convenience, $X_G$ is a discrete design space, ($G$ for grid, $X_G \subseteq \mathcal{X} \subseteq \mathbb{R}^d$) with $n_G$ number of points, initial design is defined as ($X_k,y_k$) with $k$ number of points, $X_{cand}$ ($X_{cand} \subseteq X_G$) is defined as the set of $Ncand$ candidate points. The three main criteria are:  \
\\
\begin{itemize}
\item \textit{Active Learning MacKay (ALM)}: At stage $k$ the algorithm, proposed by MacKay \citep{mackay1992}, chooses the next design point $x_{k+1}$ that maximizes the predictive variance (8) of the GP, 
\begin{equation}
x_{k+1} = \underset{x \,\in \,X_{cand}}{\arg\max} \,\, \test{s}^2(x). 
\end{equation}   
It is not as computational expensive as other sequential designs but tends to place many points in the boundaries of the design space. As the dimension size $d$ is increased, the number of boundary points grows as well. Various studies also state that the boundary points are less informative than nearby interior points \citep{Joakim2016,chaloner1995,gramacy2009,krause2008}.\
\\ 
\item \textit{Active Learning Cohn (ALC)}: At stage $k$ the algorithm, proposed by Cohn D. \citep{cohn1996},  sequentially selects the next design point $x_{t+1}$ that yields the largest reduction in predictive variance over the input space,
\begin{equation}
x_{k+1} = \underset{x \,\in \,X_{cand}}{\arg\max} \,\, \int_{\mathcal{X}} (\test{s}^2(x') - \test{s}_{k \cup x}^2(x')) \, dx',   
\end{equation} 

where $\test{s}^2(x')$ is the variance of the design point $x'$, which is already in the training data set, before observing the output at $x_{k+1}$ and $\test{s}_{k \cup x}^2(x')$ is the variance at $x'$ when the new point $x_{k+1}$ is added in the design. The integral is often approximated by a sum over a reference set, a grid of $n_{ref}$ reference points, that is,
\begin{equation}
x_{k+1} = \underset{x \,\in \,X_{cand}}{\arg\max} \,\, \frac{1}{n_{ref}} \sum_{i=1}^{n_{ref}} (\test{s}^2(x_i) - \test{s}_{k \cup x}^2(x_i))
\end{equation}   

Compared with the ALM, ALC is computationally more expensive but it performs better as it examines the effect of each point from the candidate set over the entire domain \citep{gramacy2009,seo2000}.  \
\\
\item \textit{Mutual Information for Computer Experiments (MICE)}: This algorithm, proposed by Beck and Guillas \citep{Joakim2016} is based on the information theoretic mutual information measure given by Cover and Thomas \citep{Cover2006} where the objective is to maximize the mutual information between the chosen design points and input points which have not yet been selected. It is a modified version of the mutual information criterion (MI) \citep{krause2008}. MI is described as the reduction in the uncertainty of one random vector due to the knowledge of the other. At stage $k$, the algorithm selects the next design point $x_{k+1}$ that maximizes the difference,
\begin{equation}
\max_{x \,\in\, X_{cand}} I(Y(X_k \cup x);Y(X_G \setminus (X_k \cup x))) - I(Y(X_k);Y(X_G \setminus X_k)).  
\end{equation}\
\\
For GPs, the optimization problem is written as
\begin{equation}
x_{k+1} =  \underset{x \,\in \,X_{cand}}{\arg\max} \,\, \test{s}_k^2(x) / \test{s}_{G \setminus (k \cup x)}^2(x; \tau_s^2),
\end{equation}
where $G \setminus (k \cup x)$ denotes the $X_G \setminus (X_k \cup x)$, the set (finite grid) with points that have not been selected yet and $\tau_s^2$, referred as the nugget parameter, is the extra parameter added in the correlation matrix $K$ of the GP on  $X_G\setminus(X_k \cup x)$ to achieve robustness, smoother predictions and flattening of the GP's variance \citep{Joakim2016}. In theory, the nugget parameter can take any positive value, but in practice, an ideal value is close to 1.
\end{itemize}

\subsection{GP Bandit Setting}
The strategy followed to find the optimal value in the proposed optimization scheme is to achieve a balance between exploration and exploitation. The idea is to gather more, or enough, information about the objective function by exploring the uncertain regions and then, make the best decision by exploiting (here optimizing) all the available information already known. The trade-off between exploration and exploitation has been studied in various studies within machine learning \citep{vcrepinvsek2013,ishii2002,kaelbling1996} and often seen as a multi-armed bandit problem \citep{srinivas2010,auer2002,bubeck2011,robbins1985}. \\
\\
A multi-armed bandit problem is a sequential decision making problem where at each time step of a time horizon $T$, the algorithm chooses one of the available arms (i.e. candidate points) and calculates its reward \citep{robbins1985}. When GP optimization is formulated as a bandit problem, the value of the unknown function at the chosen point $x_t$ is seen as the reward and the aim is to maximize the sum of rewards $\sum\nolimits_{t=1}^T f(x_t)$. A standard performance metric of the whole strategy of a bandit problem is the cumulative regret which measures the loss in reward due to not knowing the maximum value of $f$. To ensure that the strategy is performed well at each time step, the simple regret, $r_t$, is calculated at each iteration $t$ as $r_t = f(x^*) - f(x_t)$. Theoretical analysis of algorithms aiming to maximize the sum of rewards, under the multi-armed bandit setting, can be found in various studies \citep{srinivas2010,lai1985,azimi2010,desautels2014,contal2013,grunewalder2010}. \\
\\
The overall optimization performance is affected by whether a balance between exploration and exploitation is achieved \citep{chen2009}. To do that, acquisition functions are used that not only control the exploration-exploitation but also guide us on searching for the maximum \citep{brochu2010}. They are constructed based on the estimates obtained from the GP emulator and the current best value $f(x^*)$ achieved at each time step. A trade-off parameter, $\beta$, is added in their formulation to balance exploration and exploitation. Even if the value of the trade-off parameter has been examined in various studies, it is always left to the user \citep{Jones2001,Lizotte2008}. Among the popular choices are:
\begin{itemize}
\item \textit{Probability of Improvement (PI)}: the point with the highest PI over the best point seen so far is selected. PI is computed as
\begin{equation}
\begin{aligned}
\mbox{PI}(x) = P(f(x) \geq  f(x^*) + \beta)=\Phi \left(\frac{\test{y}(x)- f(x^*) - \beta}{\test{s}(x)}\right), 
\end{aligned}
\end{equation}
where $\Phi(\cdot)$ is the normal cumulative distribution function. 
          
\item \textit{Expected Improvement (EI)}: it considers not only the PI but also the magnitude of the improvement a point can potentially yield. The point that maximizes EI is the point that improves the unknown function the most. Under the GP, its closed form is computed as \citep{huang2006,jones1998}
\begin{equation}
\mbox{EI}(x) = (\test{y}(x) - f(x^*)) - \beta) \, \Phi \left(\frac{\test{y}(x)- f(x^*) - \beta}{\test{s}(x)}\right) + \test{s}(x) \, \phi \left(\frac{\test{y}(x)- f(\textbf{x}^*) - \beta}{\test{s}(x)}\right),
\end{equation}
where $\phi(\cdot)$ and $\Phi(\cdot)$ denote the probability density function and cumulative distribution function of the normal distribution respectively.\
\\
\item \textit{Upper Confidence Bound (UCB)}: the next point for evaluation is the one that maximizes the UCB which is computed, under the GP \citep{srinivas2010}, as 
\begin{equation}
\mbox{GP-UCB} (x) =  \underset{x \,\in \,\mathcal{X}}{\arg\max} \,\,  \test{y}(x) + \beta^{1/2} \test{s}(x). 
\end{equation} 
\end{itemize}

\section{optim-MICE: The Algorithm}
\begin{figure}[ht]
  \includegraphics[width=\textwidth]{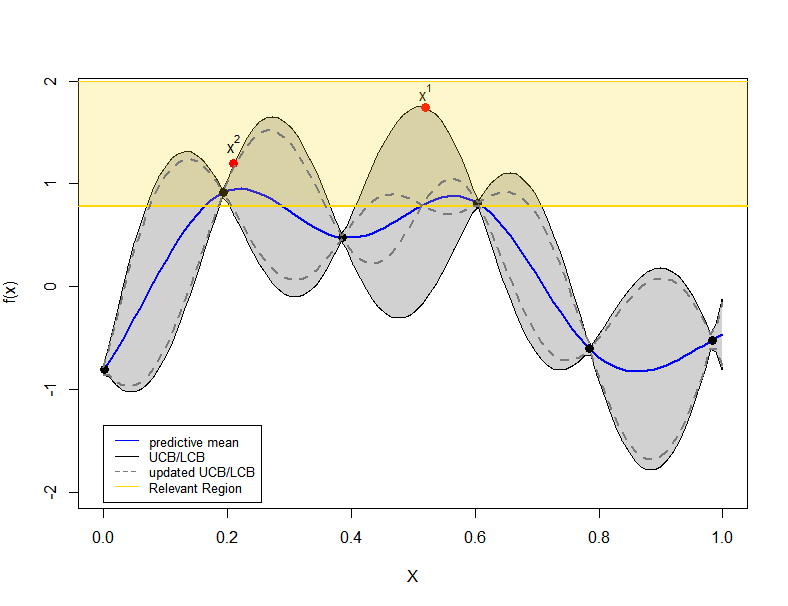}
\caption{Illustration of optim-MICE applied in $1$-dimensional test function. The grey area shows the confidence region and is bounded by $\test{f}_t^+$ and $\test{f}_t^-$. The first input point, $x^1$, is chosen based on the UCB-policy. The $\mathit{y}^{\bullet}_t$ is represented by the horizontal yellow line whereas the relevant region, $\mathfrak{R}_{t}$, is the yellow area. The black dashed lines shows the updated upper and lower bounds after having selected $x^1$. The second input point $x^2$ is chosen using the Pure Exploration strategy.}
\label{fig:1}       
\end{figure}

The newly developed sequential surrogate-based optimisation scheme follows the two-step approach of a Bayesian Optimisation framework even if Bayesian approach is not fully adapted. Firstly, the Gaussian Process regression, which is the probabilistic model, is built and then, the Upper Confidence Bound (UCB) is used, as the acquisition function, to determine at which points the black-box function will be evaluated next. At each step of the optimization process, the GP is refined by considering the new input points observed $\left\{x_t^k\right\}_{1 \leq k < K}$, which are chosen in batches of a fixed size $K$. As in GP-UCB-PE, the first point in chosen based on the UCB policy whereas the $K-1$ remaining points are chosen via the Pure Exploration strategy. \\
\\
This section gives an overview of the optimization scheme and how the new input points are chosen at each step. It also shows the technical connection between the multi-armed bandit and experimental design, as first shown in Srinivas et al. \citep{srinivas2012} and later in Contal et al. \citep{contal2013}. A simple example of the optim-MICE is illustrated in Fig. \ref{fig:1}.

\subsection{Confidence Region}
Under the GP framework, the predictive distribution at any input point $x_t$ is again a multivariate Gaussian distribution, $GP(\test{y}(x_t), \test{s}^2(x_t))$ (as given in \eqref{eq1}, \eqref{eq2}). Using this property, we can define a confidence region in which the unknown function $f$ is included with high probability. The confidence region is formed based on the well-known confidence interval. The confidence bounds are constructed as in the GP-UCB acquisition function (16) and defined as:
\begin{equation}
\begin{aligned}
\test{f}_t^{+}(x) &= \test{y}_{t}(x) + \sqrt{\beta_{t}} \test{s}_{t}(x)\\
\test{f}_t^{-}(x) &= \test{y}_{t}(x) - \sqrt{\beta_{t}} \test{s}_{t}(x),
\end{aligned}
\end{equation}
\\ 
where $f_t^+$ is the upper bound, $f_t^-$ is the lower bound, $\test{y}(x_t)$ is the predictive mean and $\test{s}^2(x_t)$ the predictive standard deviation. The width of the confidence region is regulated by the value of the trade-off parameter $\beta_t$. It controls the exploration and exploitation, namely the balance between exploring the regions with high uncertainty (regions with high predictive variance) and focusing on the supposed input point of the maximum (input point that might give the highest reward).\\
The first input point, $x_t^1$, of each batch, is chosen based upon the UCB policy and it is the one that maximizes the upper bound, or the GP-UCB acquisition function, 
\begin{equation}
x_t^1 = \underset{x \,\in \,\mathcal{X}}{\arg\max} \,\, \test{f}_t^+(x). 
\end{equation}
\subsection{Relevant Region}
Having now specified the region which contains the unknown function $f$ with high probability, a further reduction of that region is obtained where the true location of the maximum, $x^*$, of $f$ belongs with high probability. The relevant region, $\mathfrak{R}_{t}$, is defined as
\begin{equation}
\mathfrak{R}_{t} = \Big\{x \in \mathcal{X}\ | \ \test{f}_t^+(x) \geq \mathit{y}^{\bullet}_t \Big\}, 
\end{equation}\
\\
where $\mathit{y}^{\bullet}_t$ is the lower confidence bound on the maximum,  $\mathit{y}^{\bullet}_t = \test{f}_{t}^-(x^{\bullet})$ and $x_t^{\bullet} = \underset{x \,\in \,\mathcal{X}}{\arg\max} \,\, \test{f}_t^-(x)$. At every iteration $t$, only the locations that might contain the true optimum of the unknown function $f$ with high probability are kept in the relevant region, see Fig. \ref{fig:1}.

\subsection{Parallel Evaluations using MICE}
Applying the parallel strategy, we are able to choose a batch of $K$ input points at each iteration $t$. The $K-1$ remaining input points are selected via Pure Exploration. We restrict our attention to the relevant region $\mathfrak{R}_t$. The objective at this step is to maximize the information gain about the unknown function by selecting the most appropriate input points.\\
\\
In order to choose the next input point $x_t^2$, we calculate the MICE criterion (13) for all the points available for selection. The point that is selected and added in the batch is the one that maximizes the MICE criterion. For all $1 < k < K$, using a greedy strategy the new input points are selected, one by one, 
\begin{equation}
x_{t}^k =  \underset{x \,\in \,\mathfrak{R}_t}{\arg\max} \,\, \test{s}_t^{2}(x) / \test{s}_{G \setminus (t \cup x)}^{2}(x; \tau_s^2),
\end{equation}
where $\test{s}_t^2$ is the updated variance after the next input point is chosen and included in the batch. The predictive variance does not depend on the unknown function evaluation but only on the actual location of the next input point $x_t^k$. After choosing the $K-1$ input points, the uncertainty about the unknown function is reduced and the guess about the upper bound in the next iteration is improved. The overall procedure is shown in Algorithm 1.

\begin{algorithm}[H]
\caption{optim-MICE}
\label{algorithmsteps}
\begin{algorithmic}
\For{$t = 1, \ldots, T$}
 \State Compute $\test{y}_t$ and $\test{s}_t$ with Eq. (7) and (8)
 \vspace{0.2cm} 
 \State $x_t^1 = \underset{x \,\in \,\mathcal{X}}{\arg\max} \,\, \test{f}_t^+(x)$
 \vspace{0.2cm}
 \State Compute $\mathfrak{R}_t$ with Eq. (19)
 \vspace{0.2cm}
 \For {$k = 2, \ldots, K$}
  \State Compute $\test{s}_t^k$ with Eq. (8) 
  \vspace{0.2cm}
  \State $x_t^k = \underset{x \,\in \,\mathfrak{R}_t}{\arg\max} \,\, \test{s}_t^{2}(x) / \test{s}_{G \setminus (t \cup x)}^{2}(x; \tau_s^2)$
  \EndFor
\EndFor
\end{algorithmic}
\end{algorithm}

\clearpage
\section{Computational Experiments}
The empirical performance of the proposed method is compared with the GP-UCB-PE algorithm \citep{contal2013} and the q-points EI \citep{chevalier2013fast,ginsbourger2008multi}, a multi-points criterion for parallel global optimization based on the well-known EGO, on different optimization test functions. In this study, the GP-UCB-PE and the q-points EI are referred to as UCB-ALM and qEGO, respectively. The test functions have been selected to cover different input dimensions (from 2 to 6) physical properties, and shapes \citep{jamil2013literature}. Despite the fact that none of the test functions is expensive to evaluate, a meaningful study of the performance of the proposed algorithm can be conducted assuming that these functions are computationally expensive. It is expected that the proposed algorithm will behave the same on truly expensive functions as in test functions with similar surfaces.\\
\\  
The computational efficiency of the optimization algorithms is measured by calculating the cumulative regret; the loss incurred at iteration $t$ due to not knowing the input points where $f$ is maximized beforehand. Since the goal is to find the maximum of an unknown function with the lowest possible number of function evaluations, the different optimization schemes are compared in terms of the mean number of function evaluations (over multiple trials) required to get a solution with relative error $<5\%$ and $<1\%$. The relative error is given by $|f_{best}-f^*|/f^*$, provided that $f^* \neq0$, where $f_{best}$ is the best solution obtained by an algorithm and $f^*$ is the true optimum. To meet the relative error requirement of $<5\%$ and $<1\%$ the target values for each optimization test function are calculated and presented in Table \ref{tab:1}. In case where the global optimum is zero, the target values are calculated based on the range of all the possible function values ensuring that the relative error is met.

\subsection{Experimental Set-up}
In reality, the computational resources are limited, each function evaluation is costly, and the true optimum of a black-box function is unknown. The total number of function evaluations is restricted and unavoidably needs to be set in advance. So in this study, the total number of function evaluations is chosen before the optimization process starts for all the computational experiments. This number is depended by the size of the initial design drawn ($Ninit$), the number of iterations ($T$) and the batch size ($K$). Since the number of evaluations is a comparison metric for the current study, and to keep consistency, all the algorithm settings are chosen to be the same for all optimization methods. Table \ref{tab:1} gives the basic information of the test functions used in the current study and summarizes the domain and target values for each one of them. \
\\ 
\begin{table}
\centering
\caption{Test functions for the computational experiments (\textit{E})}
\label{tab:1}
\begin{tabular}{llclcrr}
\hline\noalign{\smallskip}
\multicolumn{1}{l}{\begin{tabular}[c]{@{}c@{}}Test \\ function\end{tabular}} & \multicolumn{1}{c}{Label} & Dim & \multicolumn{1}{c}{\begin{tabular}[c]{@{}c@{}}Design space\end{tabular}} & \multicolumn{1}{c}{\begin{tabular}[c]{@{}c@{}}Global \\optim\end{tabular}} & \multicolumn{1}{c}{\begin{tabular}[c]{@{}c@{}}Target\\ E \textless 1\%\end{tabular}} & \multicolumn{1}{c}{\begin{tabular}[c]{@{}c@{}}Target\\ E \textless 5\%\end{tabular}} \\ \hline
\rule{0pt}{0.4cm}Branin         & \textit{E}1  & 2   & $\left[-5,10\right]\times\left[0,15\right]$   & -0.398 & -0.402    & -0.418     \\ 
\rule{0pt}{0.4cm}Griewank       & \textit{E}2  & 2   & $\left[-600, 600\right]^2$            & 0      & -0.2     & -0.9      \\ 
\rule{0pt}{0.4cm}Himmelblau     & \textit{E}3  & 2   & $\left[-6,6\right]^2$                  & 0      & -0.2     & -1      \\ 
\rule{0pt}{0.4cm}Hosaki         & \textit{E}4  & 2   & $\left[0,10\right]^2$                  & 2.3458  & 2.3223     & 2.2285      \\ 
\rule{0pt}{0.4cm}Michalewicz    & \textit{E}5  & 2   & $\left[0,\pi\right]^2$                  & 1.8013 & 1.783     & 1.711      \\ 
\rule{0pt}{0.4cm}Sasena         & \textit{E}6  & 2   & $\left[0,5\right]^2$                   & 1.457  & 1.442     & 1.384      \\ 
\rule{0pt}{0.4cm}Six-Hump Camel & \textit{E}7  & 2   & $\left[-3,3\right]\times\left[-2,2\right]$      & 1.302  & 1.289     & 1.223       \\ 
\rule{0pt}{0.4cm}Zakharov       & \textit{E}8  & 2   & $\left[-5,10\right]^2$                 & 0      & -0.05     & -0.25       \\ 
\rule{0pt}{0.4cm}Harmann-3       & \textit{E}9  & 3   & $\left[0,1\right]^3$                   & 3.863  & 3.824     & 3.669       \\ 
\rule{0pt}{0.4cm}Rosenbrock     & \textit{E}10 & 3   & $\left[-5,10\right]^3$               & 0      & -1.8     & -9       \\ 
\rule{0pt}{0.4cm}Powell         & \textit{E}11 & 4   & $\left[-4,5\right]^4$                  & 0      & -1     & -5       \\ 
\rule{0pt}{0.4cm}Sphere         & \textit{E}12 & 4   & $\left[-5.12,5.12\right]^4$           & 0      & -0.1     & -0.5       \\ 
\rule{0pt}{0.4cm}Styblinski-Tang& \textit{E}13 & 4   & $\left[-5,5\right]^4$                  &156.664 & 155.097   & 148.831     \\ 
\rule{0pt}{0.4cm}Michalewicz    & \textit{E}14 & 5   & $\left[0,\pi\right]^5$                 & 4.688  & 4.641     & 4.453       \\ 
\rule{0pt}{0.4cm}Hartmann-6      & \textit{E}15 & 6   & $\left[0,1\right]^6$                 & 3.322  & 3.264     & 3.131       \\ 
\rule{0pt}{0.4cm}Trid           & \textit{E}16 & 6   & $\left[-36,36\right]^6$                 & 50     & 49.5      & 47.5        \\ \noalign{\smallskip}\hline
\end{tabular}
\end{table}\
\\
The initial input points are sampled using a maximin-distance design LHD. Regardless the number of dimensions, the optimization procedure for each experiment starts with an initial design of $Ninit=2$ input points. One may increase the size of $Ninit$ so as to cover better the input space with more points and possibly find the true optimum in fewer function evaluations. However, the effectiveness of using an adaptive design will not fully be achieved and this might lead to evaluate the unknown function over regions where uncertainty is low.  \\
\\
During the optimization process, $T \times K$ input points can be sequentially added in the design and therefore, in addition to the $Ninit$ runs, a further $T \times K$ evaluations can be performed. If the size of the resulting design is not restricted then choosing a bigger batch size gives the opportunity to explore more the uncertain regions without spending computational time re-estimating the GP parameters, but this increases the number of times the costly function is evaluated. In the current study, the batch size is kept fixed for all the optimization methods, regardless of the dimensions: at each time $t$, a batch of $K=5$ input points are selected. The number of iterations $T$ is changed according to the number of dimensions.\\
\\
The optimization results are affected by the size of the candidate set. Having a large number of candidate points increases the chances to end up with more input points in the Relevant Region and be closer to the true optimum but, it also increases the computational time needed to examine all the candidate points regardless of design criterion. On the other hand, with a small candidate set, we might need more function evaluations to find the true optimum and might not fully benefit from the parallel exploration as the number of candidate points available for selection in the Relevant Region might be less than the predefined batch size $K$ and therefore, the knowledge that we could obtain at each iteration for the unknown function will be minimum. \\
\\
Due to the limited computational budget, only a certain number of candidate points can be examined using the MICE criterion. As a result, to cover the whole search space and ensure that enough candidate points have been placed in the important regions, $10^4$ input points are initially sample using LHD where only a subset of them, which is randomly chosen, is examined with the MICE criterion. For convenience, the number of input points in the search space is defined as $Nsearch=10^4$ whereas the candidate points available for selection $Ncand$. At each time step $t$, a new search set of size $Nsearch=10^4$ is sampled and therefore, after the locations where $x^*$ does not belong with high probability are discarded, a new set with a number of candidate points $Ncand$ is chosen for the PE. The number of candidate points, $Ncand$, is fixed at each iteration. For computational experiments with higher dimension, the size of the candidate set is chosen to be bigger.\\
\\
For the UCB-ALM, all the input points included in the search set ($Nsearch=10^4$) are examined with the ALM criterion because its computational cost is low. Therefore, the number of candidate points available for selection is $Ncand =10^4$. The qEGO method is performed in R using the DiceKriging and DiceOptim, two packages which are built for the approximation and the optimization of black-box functions and include the q-EI criterion \citep{roustant2012dicekriging}. Specifically, at each iteration, a batch of input points is obtained by maximizing the multipoint EI criterion using a hybrid genetic algorithm. To overcome multimodality of the EI function, and keep accuracy, the default settings are kept, which does not allow us to specify the values of $Nsearch$ or $Ncand$. Furthermore, in the qEGO method, a GP model can only be fitted when the number of initial design points is bigger than the number of dimensions. As result, the $Ninit$ for the qEGO method is not kept fixed for all the computational experiments and is increased accordingly. To get an accurate measure of the performance of all the optimization schemes, 50 trials of each of these algorithms on each test function are performed where each trial uses a different random seed. Table \ref{tab:2} gives a summary of the algorithm settings used for each optimization method.\
\\
\begin{table}
\centering
\caption{Algorithm settings}
\label{tab:2}
\begin{tabular}{lcccccc}
\hline\noalign{\smallskip}
\multirow{2}{*}{\begin{tabular}[c]{@{}c@{}}Algorithm \\Settings (No.)\end{tabular}}& \multirow{2}{*}{\begin{tabular}[c]{@{}c@{}}Optimization \\ methods\end{tabular}} & \multicolumn{5}{l}{Dimensions} \\ \cline{3-7} 
\rule{0pt}{0.4cm}&  & 2D & 3D & 4D & 5D & 6D \\ \noalign{\smallskip}\hline\noalign{\smallskip}
\rule{0pt}{0.4cm}\multirow{2}{*}{\begin{tabular}[c]{@{}c@{}}Initial points  \\  \end{tabular}} & UCB-ALM, UCB-MICE                & 2  & 2 & 2 & 2 & 2  \\  
 & qEGO                          & 3  & 4 & 5 & 6 & 7  \\ 
\rule{0pt}{0.4cm}Iterations  &   & 20 &30 & 40&50 & 60 \\ 
\rule{0pt}{0.4cm}Batch size  &   & 5  & 5 & 5 & 5 & 5  \\ 
\multirow{2}{*}{\begin{tabular}[c]{@{}l@{}}Search space \\ points  \end{tabular}} & UCB-ALM, UCB-MICE & $10^4$ & $10^4$& $10^4$ & $10^4$ & $10^4$\\  
\rule{0pt}{0.4cm}& qEGO & - & - & - & - & -\\  
\rule{0pt}{0.4cm}\multirow{3}{*}{\begin{tabular}[c]{@{}l@{}}Candidate \\ points  \end{tabular}} & UCB-ALM& $10^4$ & $10^4$& $10^4$ & $10^4$ & $10^4$\\  
& UCB-MICE & 50 & 100 & 150 & 200 & 250\\  
& qEGO & - & - & - & - & - \\ \noalign{\smallskip}\hline
\end{tabular}
\end{table}

\section{Results and Discussion}
\subsection{Computational Efficiency}

\begin{figure}[!ht]
\minipage{0.45\textwidth}
\centering{\text{Best Solution}}
\vspace*{-2.8mm}
  \includegraphics[width=\linewidth]{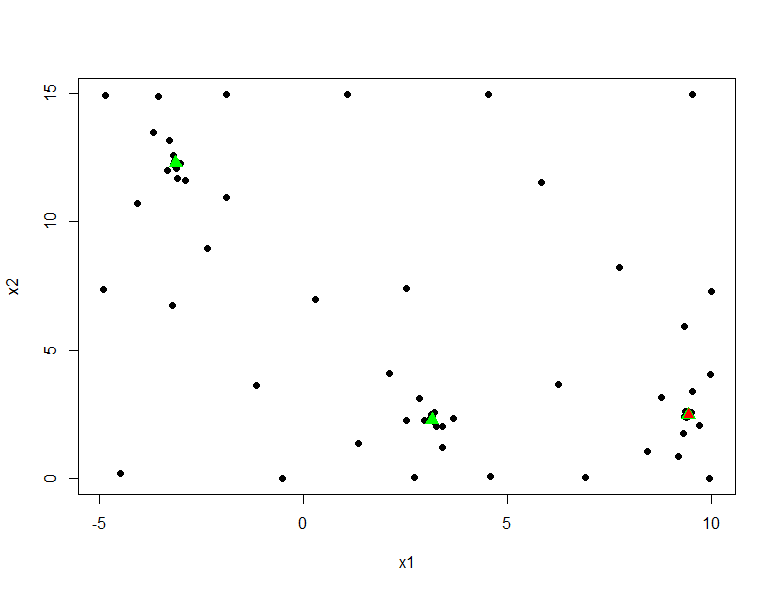}
\endminipage\hfill
\minipage{0.45\textwidth}%
\centering{\text{Worst Solution}}
\vspace*{-2.8mm}
  \includegraphics[width=\linewidth]{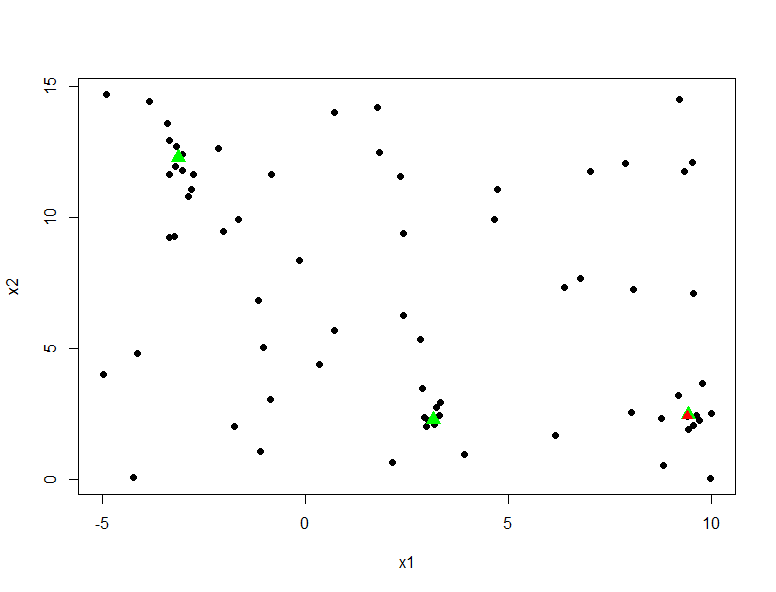}
\endminipage\\
\centering\footnotesize{UCB-ALM}\\
\minipage{0.45\textwidth}
  \includegraphics[width=\linewidth]{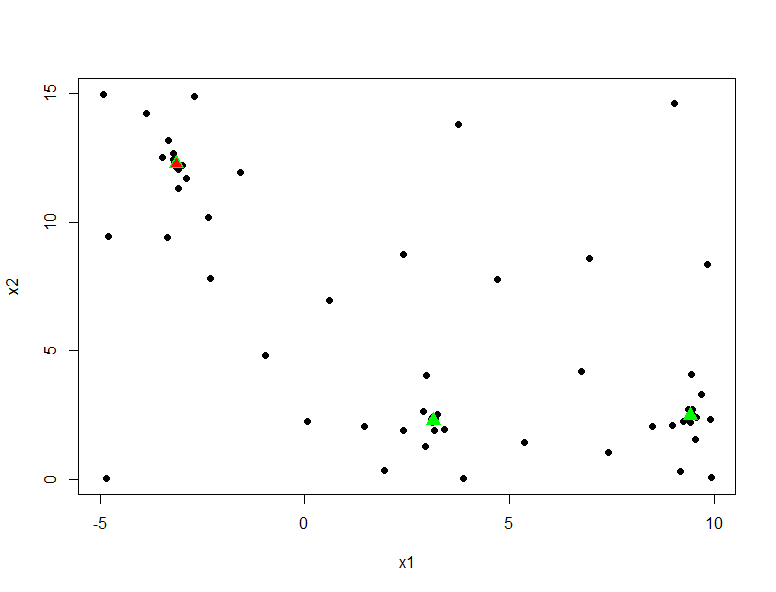}
\endminipage\hfill
\minipage{0.45\textwidth}
  \includegraphics[width=\linewidth]{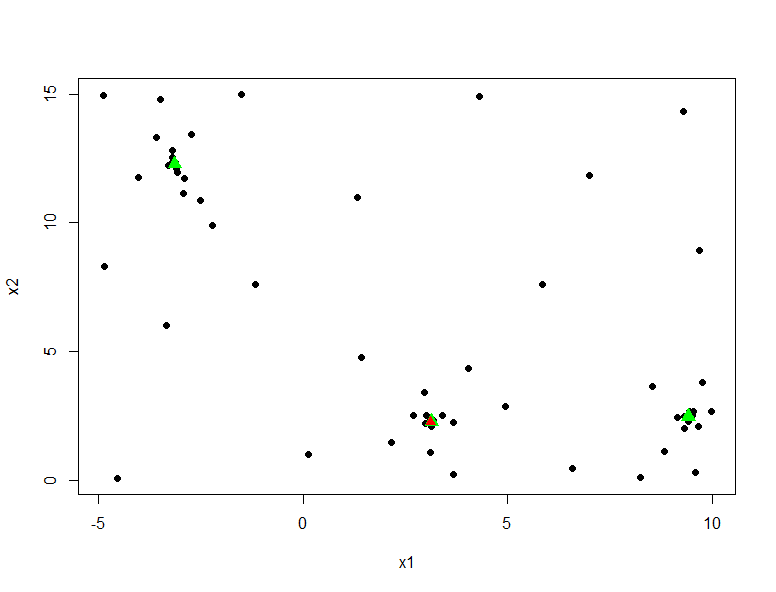}
\endminipage \hfill \\
\centering\footnotesize{UCB-MICE}\\
\minipage{0.45\textwidth}
  \includegraphics[width=\linewidth]{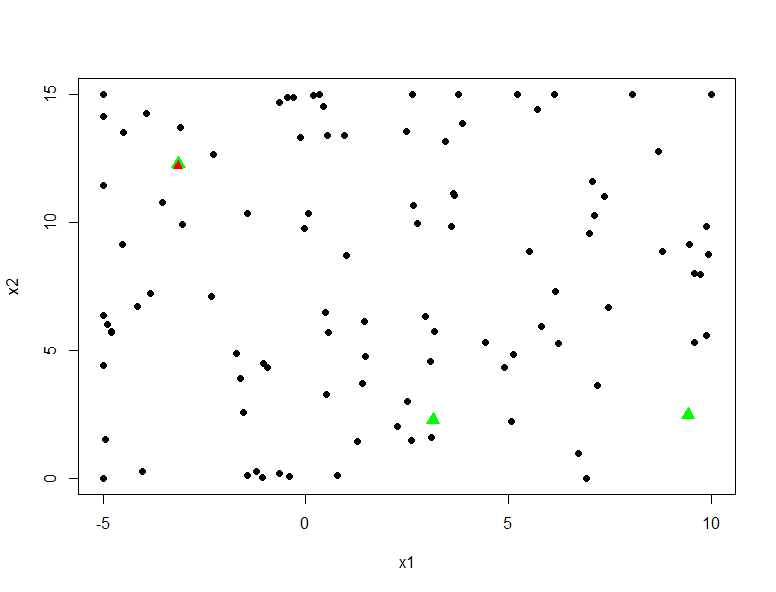}
\endminipage\hfill
\minipage{0.45\textwidth}%
  \includegraphics[width=\linewidth]{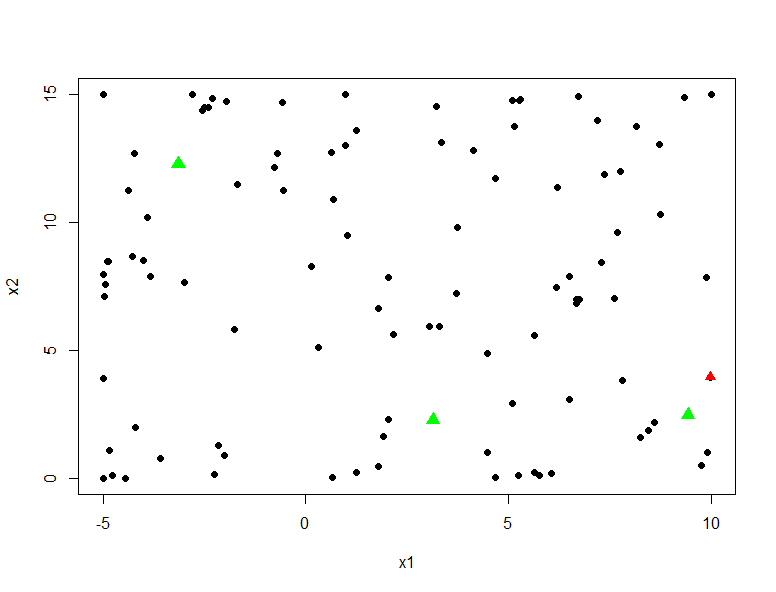}
\endminipage \hfill \\
\centering\footnotesize{qEGO}\\
\captionof{figure}{Design points selected from each optimization approach for the Branin function (\textit{E}1) according to the best (\textit{left}) and worst (\textit{right}) solution achieved among the 50 trials.}
\label{fig:3} 
\end{figure}

We compare optim-MICE with alternatives on 50 runs over 16 different computational experiments, see Fig. \ref{fig:2a}, \ref{fig:2b} and Table \ref{tab:3}. The 2-dimensional experiments examined are W-shaped (\textit{E}4, \textit{E}6), steep ridges (\textit{E}5), valley-shaped (\textit{E}7), plate-shaped (\textit{E}8) with either one global maximum or more (\textit{E}1) and sometimes with many local maxima (\textit{E}2). Regardless of the surface and the complexity of the function, UCB-MICE and UCB-ALM are doing definitely better than qEGO as all the best solutions obtained in each trial (summarized in box-plots in Fig. \ref{fig:2a} \& \ref{fig:2b}) are closer to the true optimum. For \textit{E}1, \textit{E}3, \textit{E}5, \textit{E}6 and \textit{E}7, the target values and the true optimum in all the 50 trials performed are fully achieved (Table \ref{tab:3}). For \textit{E}2, \textit{E}4 and \textit{E}8 not all the trials are successful as some of them do not get a solution within $1\%$ and/or $5\%$ of the global maximum. However, the mean best solution, for both ALM- and MICE-based algorithms, always give a value close to the true optimum. On the other hand, considering the spread of the solutions obtained in qEGO and the small number of the successful trials, the behaviour of algorithm can be unstable and the optimum achieved can be far away from the targets. A substantial difference between UCB-ALM and UCB-MICE can be noted on the mean function evaluations required to meet the target values. Using UCB-MICE, the target values can be achieved faster, with fewer function evaluations, which imparts confidence in the computational efficiency of MICE. \\
\\
In the higher dimensional experiments, the computational complexity increase and the target values are harder to achieve. Again, using UCB-MICE surpasses the other two approaches, as for all the experiments (\textit{E}9-\textit{E}16) the number of successful trials is greater. To get a solution with a relative error $<5\%$ and $<1\%$, with optim-MICE, requires less function evaluations and therefore less computational resources. For example, a solution with a relative error $<1\%$ for the \textit{E}9 and \textit{E}15 is achieved by evaluating the function 10 and 14 fewer times, compared to the other algorithms, respectively (Table \ref{tab:3}). \\
\\
The summary results for the experiments \textit{E}9, \textit{E}13, \textit{E}14 and \textit{E}15 presented in Fig. \ref{fig:2b} suggest that optim-MICE outperforms competing methods. The exact shape of \textit{E}9 and \textit{E}15 (Hartmann function) is unknown but it is well-known that it is a relatively smooth function with very few modes. The mean best solution for the 3- and 6-dimensional cases indicates UCB-ALM and UCB-MICE are doing better than qEGO. A difficult test case is \textit{E}14 (Michalewicz 5d) because part of the surface is plateau, which often hampers the search process of the optimization algorithm as these areas do not offer any information, and has steep ridges. The complexity of the function does not seem to be a problem in the 2-dimensional case (\textit{E}5) but here, all the algorithms struggle to achieve the target values and find the global maximum. Comparing the three methods, optim-MICE does significantly better than the alternatives as, in most of the 50 trials performed, the best solution achieved is closer to the true optimum.\\
\\
\begin{subfigures}
\begin{figure}[ht]
\vspace{-1cm}
\centering
\minipage{0.335\textheight}
  \includegraphics[width=\textwidth]{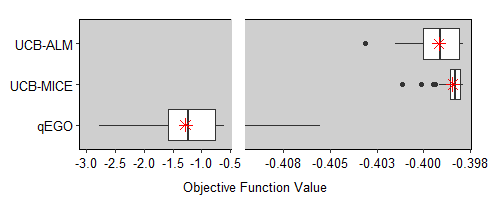}
\centering\footnotesize{Branin2D (\textit{E}1)}
\vspace{0.3cm}
\endminipage\hfill
\minipage{0.335\textheight}
  \includegraphics[width=\textwidth]{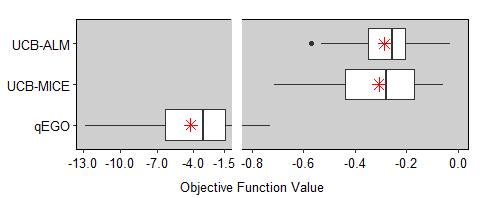}
 \centering\footnotesize{Griewank2D (\textit{E}2)}
 \vspace{0.3cm}
\endminipage\hfill\\
\minipage{0.335\textheight}
  \includegraphics[width=\textwidth]{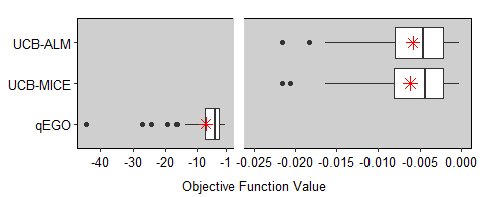}
 \centering\footnotesize{Himmelblau2D (\textit{E}3)}
 \vspace{0.3cm}
\endminipage\hfill
\minipage{0.335\textheight}
  \includegraphics[width=\textwidth]{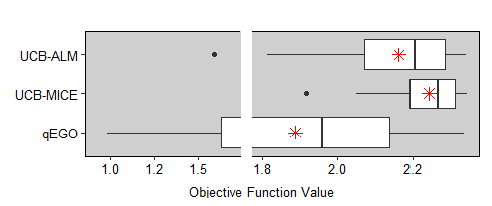}
   \centering\footnotesize{Hosaki2D (\textit{E}4)}
   \vspace{0.3cm}
\endminipage\\
\minipage{0.335\textheight}
  \includegraphics[width=\textwidth]{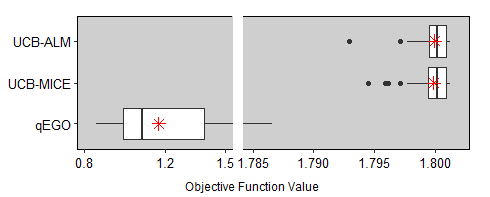}
     \centering\footnotesize{Michalewicz2D (\textit{E}5)}
     \vspace{0.3cm}
\endminipage\hfill
\minipage{0.335\textheight}
  \includegraphics[width=\textwidth]{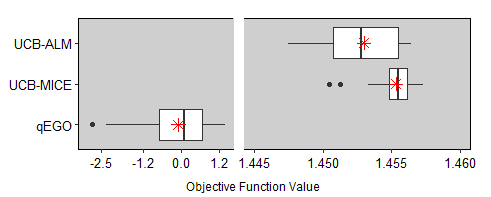}
     \centering\footnotesize{Sasena2D (\textit{E}6)}
     \vspace{0.3cm}
\endminipage\hfill\\
\minipage{0.335\textheight}
  \includegraphics[width=\textwidth]{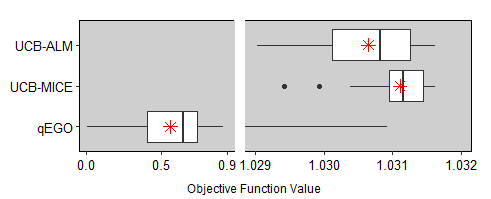}
     \centering\footnotesize{Six-Hump2D (\textit{E}7)}
     \vspace{0.3cm}
\endminipage\hfill
\minipage{0.335\textheight}
  \includegraphics[width=\textwidth]{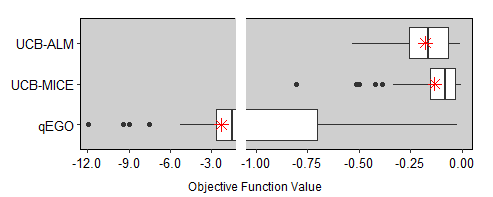}
   \centering\footnotesize{Zakharov2D (\textit{E}8)}
     \vspace{0.3cm}
\endminipage\hfill
\captionof{figure}{Summary of the best solution achieved in the 50 trials in box-plots with a gap in the range of function values. Red star shows the mean best solution.}
\label{fig:2a} 
\end{figure}\ 
\\
The global optimum of \textit{E}10 (Rosenbrock) is located in a long narrow valley which makes the exploration process of an optimization algorithm even slower. Using either the ALM or MICE criterion, it is ensured to achieve the target values and a solution closer to the true optimum. As in \textit{E}7, which is also a valley-shaped function, so too in \textit{E}10, the function values obtained with qEGO are far away from the true optimum indicating its difficulty to find the uncertain region. For most of the global optimization methods relied on heuristic techniques, \textit{E}11 (Powell) is an easy problem \citep{steihaug2013global} whereas for other methods, including the proposed algorithm, it is a challenging test case. Considering the summary results in Fig. \ref{fig:2b}, UCB-MICE and qEGO are significantly better than UCB-ALM. What attracts the attention here is that with qEGO the mean function evaluations required to get a solution within $5\%$ of the true optimum is much lower, and the number of successful trials is higher, than the alternatives but none of the trials performed gets a solution with a relative error $<1\%$. On the other hand, with UCB-MICE, the best solution achieved is even closer to the true optimum. \\ 
\\
Both functions, \textit{E}12 (Sphere 4D) and \textit{E}16 (Trid 6D), have a bowl-shaped surface but, based on their 2-dimensional illustrations, the uncertain region of \textit{E}12 is in the middle of the surface while Trid's global maximum is positioned towards the edge. In contrast with UCB-ALM, UCB-MICE spots the uncertain region of \textit{E}12 faster and the mean function evaluations required to achieve the target values is smaller. This is also in line with the structure of the MICE criterion which, at each time step, forces us to choose a new point that yields the least uncertainty between itself and the unselected input points. This point is usually 'central' with respect to points that have not been selected yet \citep{krause2008}. As ALM tends to place many points in the boundaries of the design space, it is expected to find the uncertain region faster and perform better in \textit{E}16. The number of trials that found a function value within $5\%$ of the true optimum is higher than in UCB-MICE but none of the trials got a solution with a relative error $<1\%$ (Table \ref{tab:3}). Considering the range of the best solutions achieved in the 50 trials (Fig. \ref{fig:2b}), it can be stated that the overall performance of optim-MICE is better.

\begin{figure}[!ht]
\centering
\minipage{0.335\textheight}
  \includegraphics[width=\textwidth]{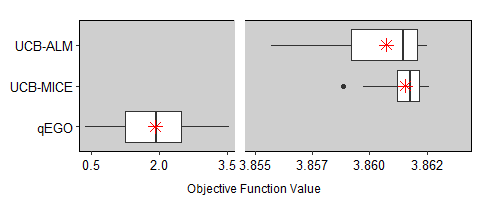}
       \centering\footnotesize{Hartmann3D (\textit{E}9)}
       \vspace{0.3cm}
\endminipage\hfill
\minipage{0.335\textheight}
  \includegraphics[width=\textwidth]{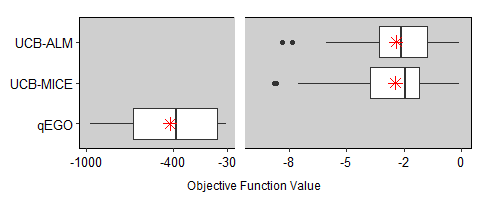}
       \centering\footnotesize{Rosenbrock3D (\textit{E}10)}
       \vspace{0.3cm}
\endminipage\hfill\\
\minipage{0.335\textheight}
  \includegraphics[width=\textwidth]{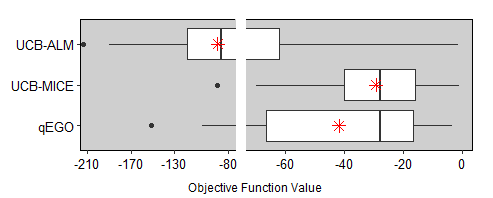}
         \centering\footnotesize{Powell4D (\textit{E}11)}
         \vspace{0.3cm}
\endminipage\hfill
\minipage{0.335\textheight}
  \includegraphics[width=\textwidth]{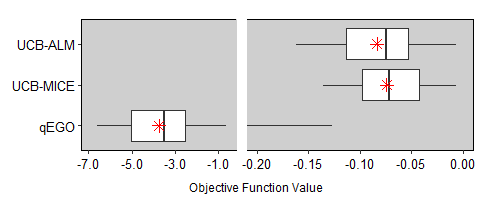}
           \centering\footnotesize{Sphere4D (\textit{E}12)}
           \vspace{0.3cm}
\endminipage\hfill\\
\minipage{0.335\textheight}
  \includegraphics[width=\textwidth]{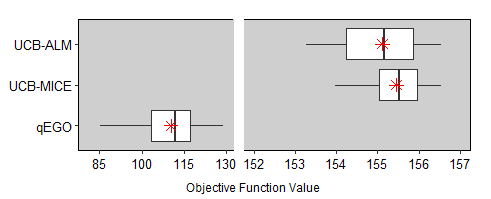}
       \centering\footnotesize{Styblinski4D (\textit{E}13)}
       \vspace{0.3cm}
\endminipage\hfill
\minipage{0.335\textheight}
  \includegraphics[width=\textwidth]{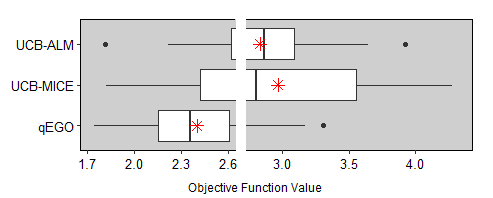}
       \centering\footnotesize{Michalewicz5D (\textit{E}14)}\\
       \vspace{0.3cm}
\endminipage\hfill\\
\minipage{0.335\textheight}
  \includegraphics[width=\textwidth]{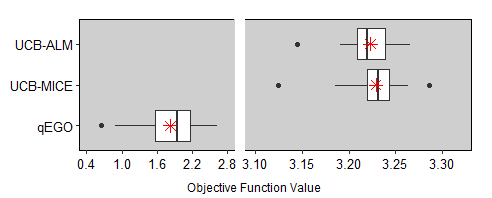}
         \centering\footnotesize{Hartmann6D (\textit{E}15)}
         \vspace{0.3cm}
\endminipage\hfill
\minipage{0.335\textheight}
  \includegraphics[width=\textwidth]{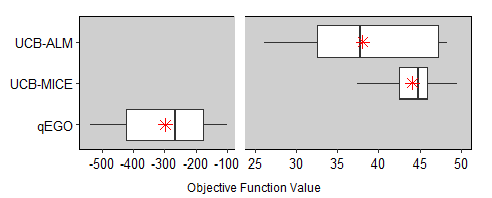}
           \centering\footnotesize{Trid6D (\textit{E}16)}
\endminipage\hfill
\captionof{figure}{Summary of the best solution achieved in the 50 trials in box-plots with a gap in the range of function values. Red star shows the mean best solution.}
\label{fig:2b} 
\end{figure}
\end{subfigures}\ 
\\
Fig. \ref{fig:3} shows all the design points chosen for \textit{E}1 (Branin), from each algorithm, of the trial with the best (left) and worst (right) solution. The Branin function is a 2-dimensional case with three global maxima therefore, with three regions where uncertainty in higher. Choosing the design points based on the MICE criterion proves to be advantageous as in both trials, best and worst, the uncertain regions are fully explored without spending computational time on unnecessary areas. Although with the ALM criterion the uncertain regions are also successfully spotted on both trials, an amount of computational time is spent exploring unimportant regions and evaluating the objective function on locations far away from the true optimum. It also tends to push a number of design points near the boundaries which might make it attractive for cases where the optimum is in the edges. Considering the input points chosen with qEGO, the visited locations are spread around the entire design space. As in the best trial, so too to the worst trial, it lacks the ability to identify the important regions, and stick to them, which results in a bad design set and often the true optimum is not found within the predefined time horizon. The information gained about the objective function from the chosen points is limited, and a valuable computational time is wasted on evaluating it on unhelpful locations. A possible reason that might affect the overall performance of qEGO is the small number of initial points: at the initial stage of the optimization, the knowledge is minimal and the right direction is difficult to be found.\ 
\\
\begin{table}
\centering
\begin{minipage}{\textwidth}
\caption{Mean function evaluations required to get a solution with a relative error $< 1\%$ and $< 5\%$. Brackets show the number of successful trials, out of the 50 performed, which achieved the target values.}
\label{tab:3}
\begin{tabular}{|lrrr|rrr|rrr|rrr|rrr|}
\hline\noalign{\smallskip}
\begin{tabular}[c]{@{}l@{}}E \\ \end{tabular} & \multicolumn{3}{c}{\begin{tabular}[c]{@{}c@{}}Mean Function \\ Evaluations $E<1\%$\end{tabular}} & \multicolumn{3}{c|}{\begin{tabular}[c]{@{}c@{}}Mean Function \\ Evaluations $E<5\%$\end{tabular}} \\ \cline{2-7} 
\rule{0pt}{0.4cm} & \multicolumn{1}{c}{UCB-ALM} & \multicolumn{1}{c}{optim-MICE} & \multicolumn{1}{c}{qEGO} & \multicolumn{1}{c}{UCB-ALM} & \multicolumn{1}{c}{optim-MICE} & \multicolumn{1}{c|}{qEGO} \\ \hline
\rule{0pt}{0.4cm}\textit{E}1 &    52(50) & \textbf{49(50)} & 100+(0) & 41(50) & \textbf{39(50)} & 100+(0) \\ 
\rule{0pt}{0.4cm}\textit{E}2 &    64(12) & \textbf{50(13)} & 100+(0) & 23(50) & \textbf{21(50)} & 8(6) \\
\rule{0pt}{0.4cm}\textit{E}3 &    44(50) & \textbf{39(50)} & 71(3) & 32(50) & \textbf{30(50)} & 73(11) \\ 
\rule{0pt}{0.4cm}\textit{E}4 &    89(4)  & 71(9)  & \textbf{49(4)} & 61(29) & 57(41) & \textbf{39(7)} \\ 
\rule{0pt}{0.4cm}\textit{E}5 &    59(50) & \textbf{58(50)} & 67(3) & 55(50) & \textbf{53(50)} & 57(5) \\ 
\rule{0pt}{0.4cm}\textit{E}6 &    75(50) & \textbf{70(50)} & 100+(0) & \textbf{52(50)} & 57(50) & 38(3) \\ 
\rule{0pt}{0.4cm}\textit{E}7 &    57(50) & \textbf{51(50)} & 79(5) & 44(50) & \textbf{40(50)} & 47(11) \\ 
\rule{0pt}{0.4cm}\textit{E}8 &    79(11) & 78(26) & \textbf{45(4)} & 74(37) & 67(42) & \textbf{41(11)} \\ 
\rule{0pt}{0.4cm}\textit{E}9 &    45(50) & \textbf{35(50)} &  150+(0) & \textbf{28(50)} & 35(50) & 150+(0) \\ 
\rule{0pt}{0.4cm}\textit{E}10 &  120(21) & \textbf{116(29)} & 150+(0)  &104(50) & \textbf{100(50)} & 150+(0) \\ 
\rule{0pt}{0.4cm}\textit{E}11 &     200+(0) & \textbf{197(13)} & 200+(0)  & 191(7) & 183(17) & \textbf{151(23)} \\ 
\rule{0pt}{0.4cm}\textit{E}12 &   52(31) & \textbf{45(34)} & 200+(0)  & 29(50) & \textbf{26(50)} & 167(15) \\ 
\rule{0pt}{0.4cm}\textit{E}13 &  103(36) & \textbf{97(39)} & 200+(0)  & 76(50) & \textbf{70(50)} & 200+(0) \\ 
 \rule{0pt}{0.4cm}\textit{E}14&     250+(0) & 250+(0)   & 250+(0)  & 250+(0)   & 250+(0) & 250+(0) \\ 
\rule{0pt}{0.4cm}\textit{E}15 &     193(3) & \textbf{179(9)}   & 300+(0)  & 88(50) & \textbf{82(49)} & 300+(0) \\ 
\rule{0pt}{0.4cm}\textit{E}16 &  300+(0) & \textbf{119(8)} & 300+(0) & 107(14) & \textbf{102(11)} & 300+(0) \\  \noalign{\smallskip}\hline
\end{tabular}\footnotetext{Numbers in bold indicate the best result achieved among the three optimization schemes.}
\end{minipage}
\end{table}\ 
\\
\
\\
\ 
\\
The way how qEGO explores the input space looks to be beneficial when the true optimum is not isolated in a small region of the search space. Such examples are \textit{E}4 and \textit{E}8, where the uncertain region covers a huge area and is not interrupted by ridges or drops. Despite the fact that qEGO could not achieve the target values in most of the trials performed in \textit{E}4 or \textit{E}8, and the mean best solution is far from the true optimum (Fig. \ref{fig:2a}), for the ones that perform well, the average function evaluations required to a get a solution with a relative error $<5\%$ and $<1\%$ is much lower compared with the other methods. By slightly increasing the number of initial design, and the corresponding initial computational cost, qEGO could possibly perform better in some particular experiments.\\
\\
The computational efficiency of the three algorithms is further examined by calculating the simple regret. Fig. \ref{fig:4} shows the evolution of the mean simple regret for nine of the experiments. The small plots are zoomed around the most interesting part. The mean simple regret is calculated based on the true optimum and the mean current solution taken of the 50 runs.\\
\\
In more details, with ALM- and MICE-based criterion, the regret in \textit{E}1 converges to zero after the same number of function evaluations whereas, with qEGO, this never happens in the specified time horizon. A difficult case for all the algorithms, as it is mentioned earlier, is when the objective function has many local maxima. Such a case is the \textit{E}2 where, with qEGO, even if a lot of function evaluations are performed, the regret is not converging to zero. In contrast, the decay of regret is done faster in UCB-ALM and UCB-MICE but, to converge to zero and get a solution literally close to the global maximum both need a lot more function evaluations. The efficiency of the search process followed by the MICE-based algorithm is also noted in \textit{E}7, \textit{E}8, \textit{E}13 and \textit{E}16 as the convergence is achieved in fewer function evaluations than in alternatives. In terms of qEGO, except in \textit{E}9 where it seems to struggle a lot, in all the other experiments the regret decreases, but slowly, and does not converge to zero within the predefined time period common to all methods. As it is expected, despite the improvement seen in \textit{E}14 during the optimization procedure, the complexity level of the function affects the algorithm's search process. To achieve convergence with any of the three methods, more function evaluations would be required, and tuning the algorithm settings would be also helpful. Overall, considering the fast decay of the regret, it can be stated that UCB-MICE finds the uncertain region and gets a solution close to the true optimum in less computational time, regardless of dimensionality and complexity of the function. 

\subsection{Tunning Settings}
The main settings of optim-MICE are the number of iterations (\textit{T}), the batch size (\textit{K}), the number of input points in the search space available for selection (\textit{Nsearch}) and the number of candidate points ready to be examined with MICE (\textit{Ncand}). To understand their effect on the proposed algorithm, a simple sensitivity analysis is performed on four of the experiments: Hosaki 2D (\textit{E}4), Sasena 2D (\textit{E}6), Rosenbrock 3D (\textit{E}10) and Hartmann 6D (\textit{E}15). Specifically, one setting is varied while others remain fixed. In total, 12 different scenarios are presented and compared over each test function. The base scenario has the same algorithm settings as in the previous section (Table \ref{tab:2}). The effect of the main settings on the overall performance of the algorithm, and whether its computational efficiency is improved, are also examined in a scaled version of $f(x)$. At two different levels, Hosaki 2D (\textit{E}4) and Rosenbrock 3D (\textit{E}10) are both scaled vertically and horizontally, by varying one algorithmic setting at a time, and compared with the non-scaled version. As before, 50 trials performed for each scenario.\\
\\
\begin{landscape}
 \begin{figure}[ht]
\minipage{0.43\textwidth}
\vspace*{-4.8mm}
  \includegraphics[width=\linewidth]{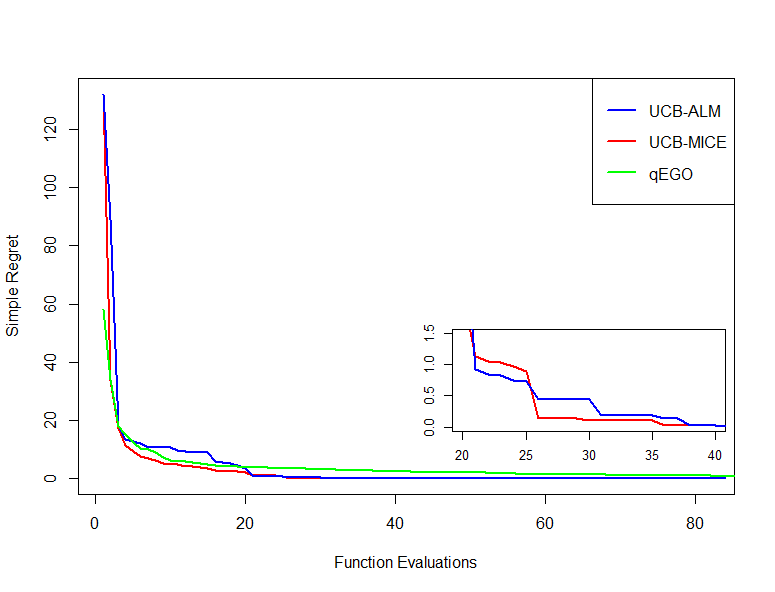}
\centering\small{\text{Branin2D (\textit{E}1)}}
\endminipage\hfill
\minipage{0.43\textwidth}
\vspace*{-4.8mm}
  \includegraphics[width=\linewidth]{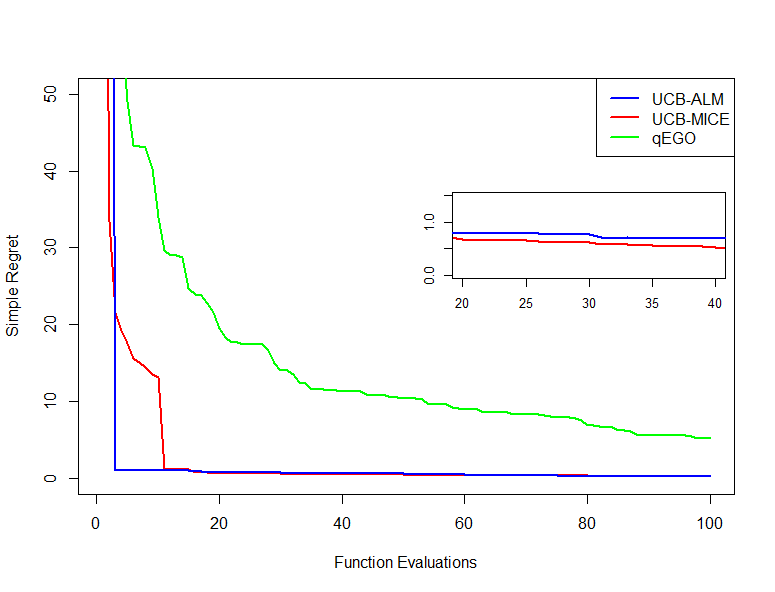}
 \centering\small{\text{Griewank2D (\textit{E}2)}}
\endminipage\hfill
\minipage{0.43\textwidth}%
\vspace*{-4.8mm}
  \includegraphics[width=\linewidth]{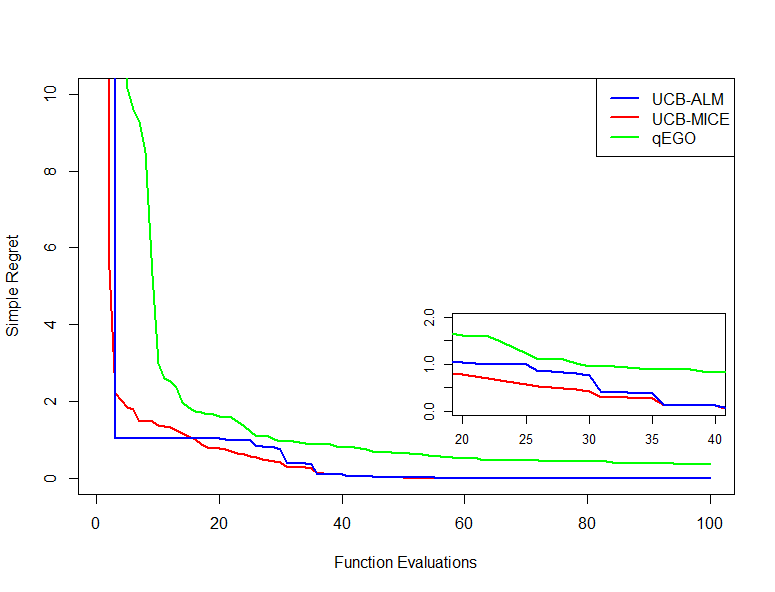}
   \centering\small{\text{Six-Hump2D (\textit{E}7)}}
\endminipage\\
\minipage{0.43\textwidth}
  \includegraphics[width=\linewidth]{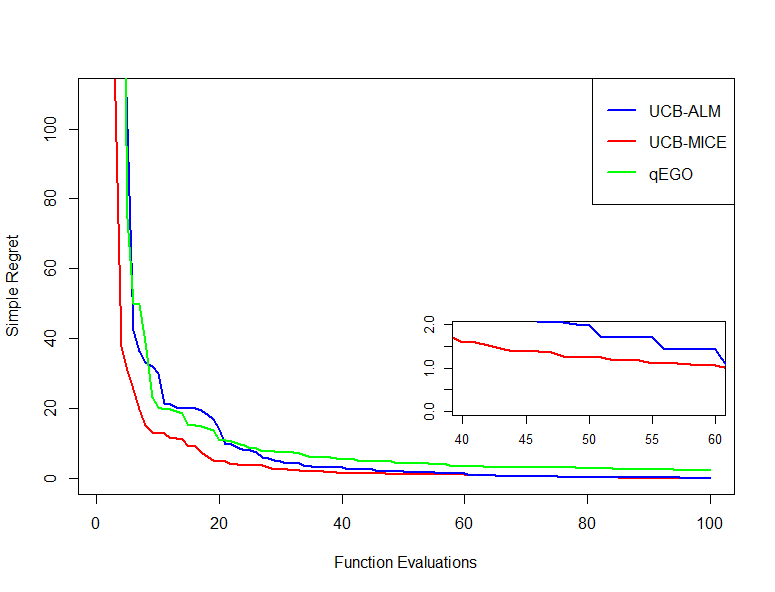}
     \centering\small{\text{Zakharov2D (\textit{E}8)}}
\endminipage\hfill
\minipage{0.43\textwidth}
  \includegraphics[width=\linewidth]{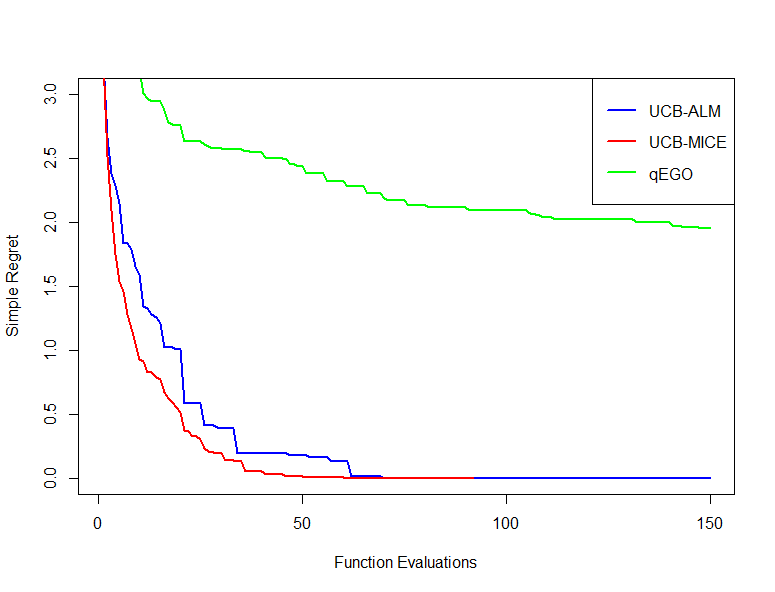}
     \centering\small{\text{Hartman3D (\textit{E}9)}}
\endminipage\hfill
\minipage{0.43\textwidth}
  \includegraphics[width=\linewidth]{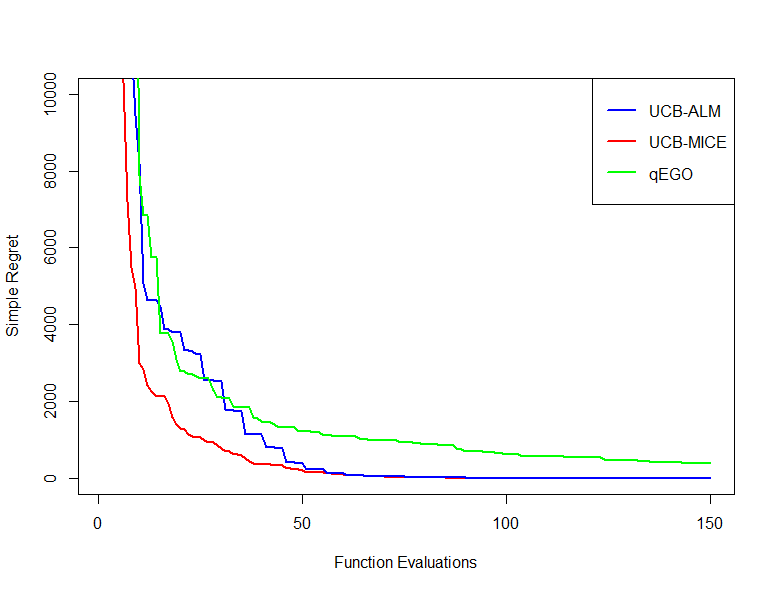}
     \centering\small{\text{Rosenbrock3D (\textit{E}10)}}
\endminipage\\
\minipage{0.43\textwidth}
  \includegraphics[width=\linewidth]{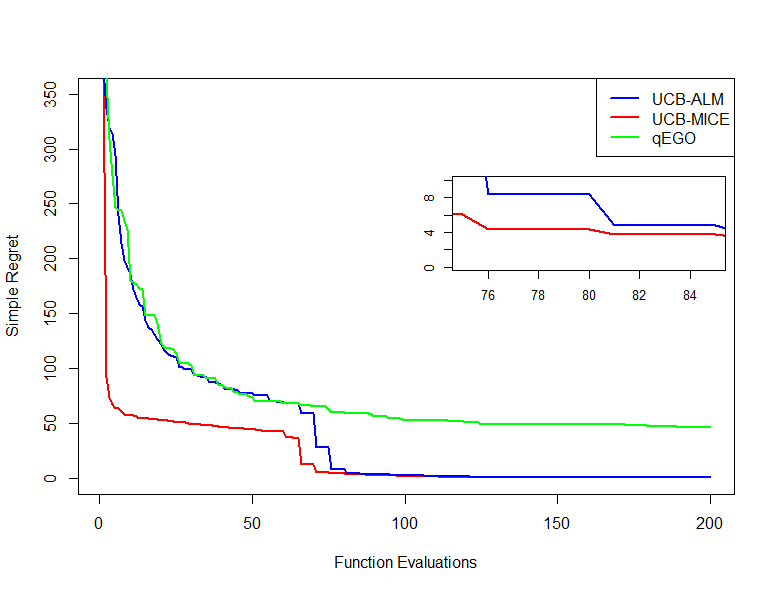}
       \centering\small{\text{Styblinski4D (\textit{E}13)}}
\endminipage\hfill
\minipage{0.43\textwidth}
  \includegraphics[width=\linewidth]{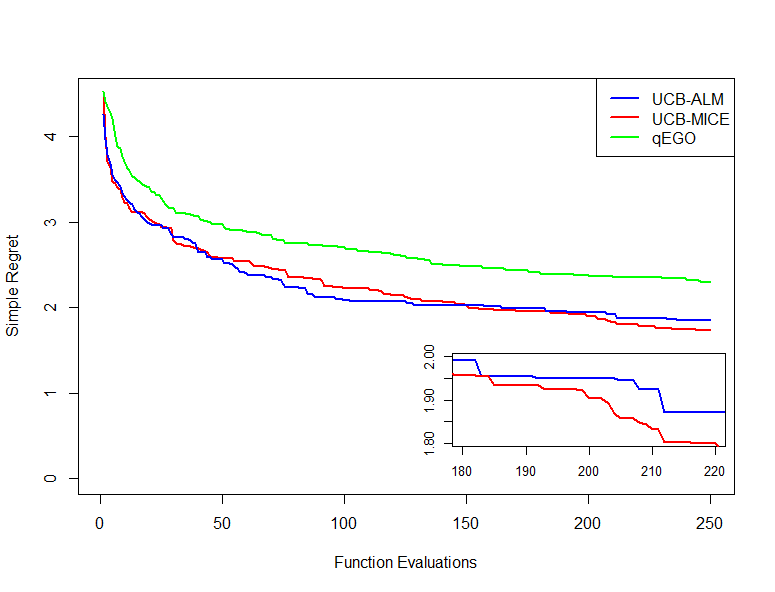}
         \centering\small{\text{Michalewicz5D (\textit{E}14)}}
\endminipage\hfill
\minipage{0.43\textwidth}
  \includegraphics[width=\linewidth]{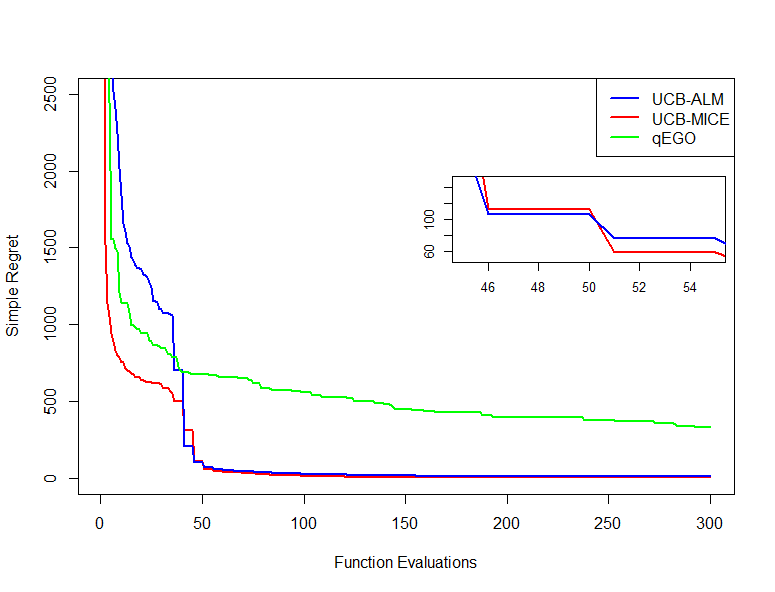}
           \centering\small{\text{Trid6D (\textit{E}16)}}
\endminipage\hfill
\captionof{figure}{Comparison of the mean simple regret with respect to the total function evaluations performed during the optimization process. Small plots show a zoomed part of the decay of the regret.}
\label{fig:4} 
\end{figure}
\end{landscape} \
\\

\begin{landscape}
\begin{figure}[ht]
\centering
\includegraphics[height=280pt, width=570pt]{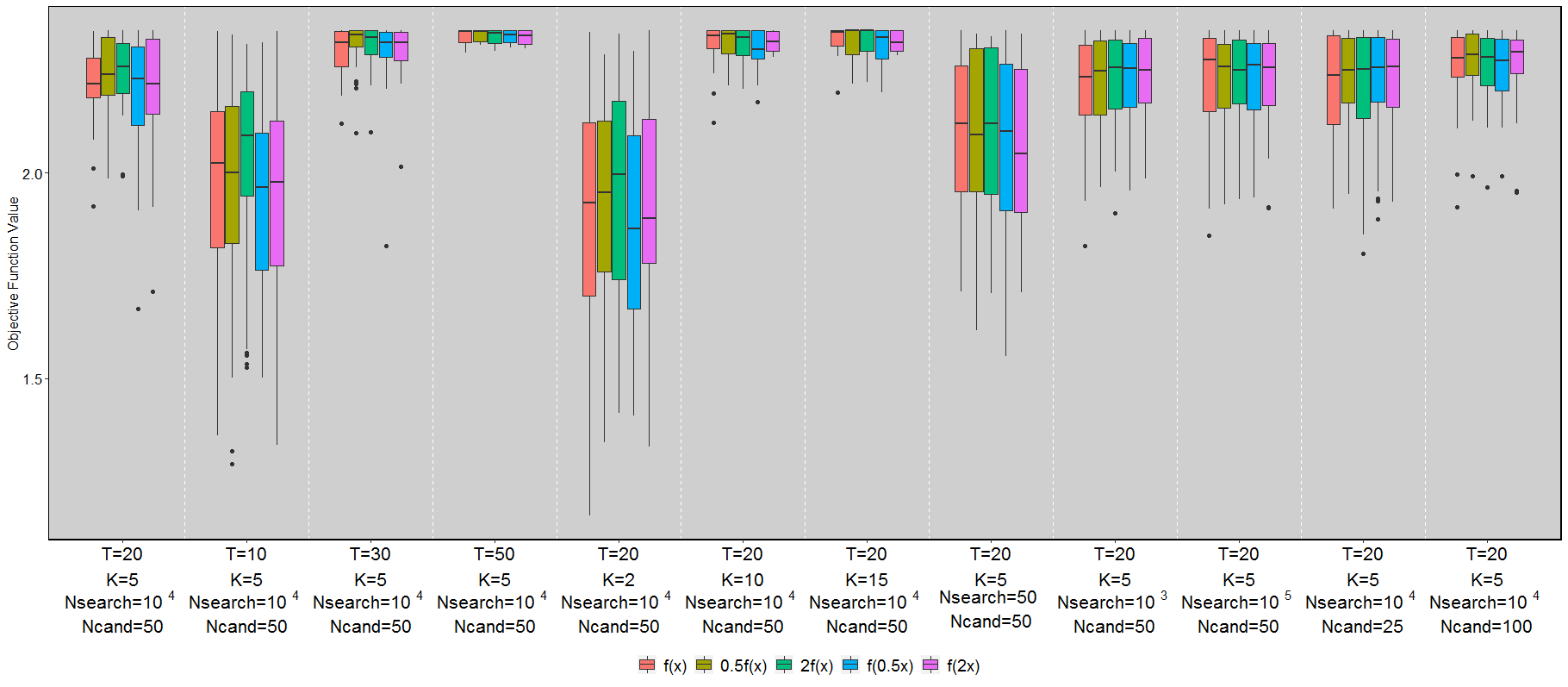}
\caption{Summary of the best solution achieved in the 50 trials in box-plots for the scaled-versions of Hosaki 2D (\textit{E}4)}
\label{fig:5} 
\end{figure}
\end{landscape}

\begin{landscape}
\begin{figure}[ht] 
\centering
\includegraphics[height=280pt, width=570pt]{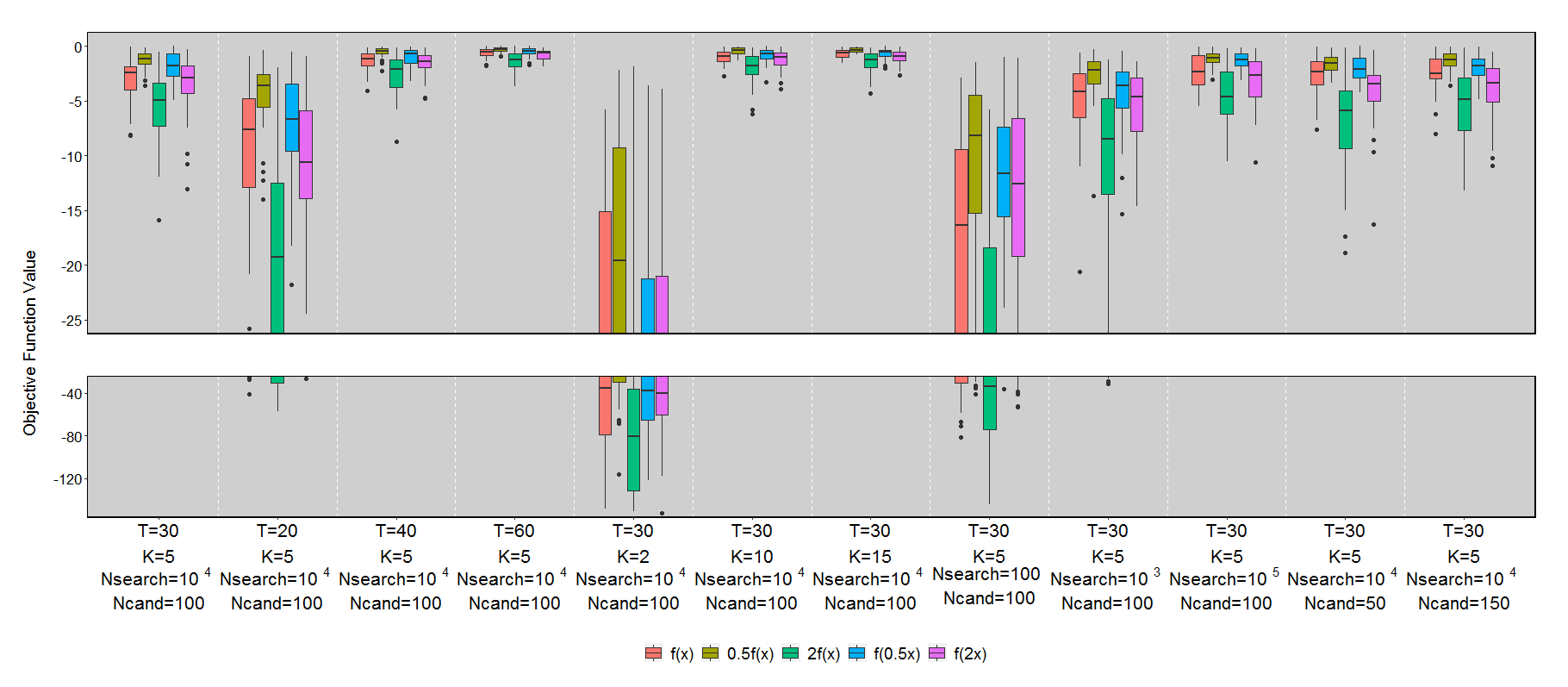}
\caption{Summary of the best solution achieved in the 50 trials in box-plots  with a gap in the range of function values for the scaled-versions of Rosenbrock 3D (\textit{E}10)}
\label{fig:6} 
\end{figure}
\end{landscape}

\begin{table}
\begin{minipage}{\textwidth}
\caption{Tunning Algorithm Settings: Scaling Hosaki 2D (\textit{E}4). Brackets show the number of successful trials, out of the 50 performed, which achieved a solution with a relative error $<5\%$.}
\label{tab:4}
\begin{tabular}{|ccccccccc|}
\hline\noalign{\smallskip}
 \multirow{5}{*}{T} & \multirow{5}{*}{K} & \multirow{5}{*}{Nsearch} & \multirow{5}{*}{Ncand} & \multicolumn{5}{c|}{Mean Function Evaluations (E \textless 5\%)} \\ \cline{5-9} 
\rule{0pt}{0.35cm} &  &  &  & \multicolumn{1}{c}{\multirow{4}{*}{\begin{tabular}[c]{@{}c@{}}Non-\\ Scaled \\ Version\end{tabular}}} & \multicolumn{4}{c|}{Scaled Versions} \\ \cline{6-9} 
 &  &  &  & \multicolumn{1}{c}{} & \multicolumn{2}{c}{\rule{0pt}{0.35cm}Vertical} & \multicolumn{2}{c|}{Horizontal} \\ \cline{6-9} 
\rule{0pt}{0.5cm} &  &  &  & \multicolumn{1}{c}{} & $0.5f(x)$ & $2f(x)$ & $f(0.5x)$ & $f(2x)$ \\ \hline
20 & 5 & $10^4$ & 50 & 57(41) & \cellcolor[HTML]{C0C0C0}55(44) & 57(44) & 60(41) & 59(42) \\ 
\textbf{10} & 5 & $10^4$ & 50 & 28(12) & \cellcolor[HTML]{C0C0C0}22(12) & 22(10) & 29(8) & 29(7) \\ 
\textbf{30} & 5 & $10^4$ & 50 & 84(47) & \cellcolor[HTML]{C0C0C0}80(45) & 81(47) & 82(44) & 83(46) \\ 
\textbf{50} & 5 & $10^4$ & 50 & 83(50) & \cellcolor[HTML]{C0C0C0}75(50) & 77(50) & 84(50) & 85(50) \\ 
20 & \textbf{2} & $10^4$ & 50 & 30(6) & \cellcolor[HTML]{C0C0C0}18(5) & 20(6) & 21(4) & 23(8) \\ 
20 & \textbf{10} & $10^4$ & 50 & 98(48) & \cellcolor[HTML]{C0C0C0}94(48) & 97(49) & 104(48) & 107(50) \\ 
20 & \textbf{15} & $10^4$ & 50 & 92(50) & \cellcolor[HTML]{C0C0C0}89(50) & 92(50) & 102(50) & 105(50) \\ 
20 & 5 & \textbf{50} & 50 & 58(26) & \cellcolor[HTML]{C0C0C0}58(35) & 58(36) & 58(33) & 58(33) \\
20 & 5 & \bm{$10^3$} & 50 & 58(30) & \cellcolor[HTML]{C0C0C0}57(39) & 57(32) & 61(32) & 63(31) \\ 
20 & 5 & \bm{$10^5$} & 50 & 54(41) & \cellcolor[HTML]{C0C0C0}51(43) & 53(40) & 55(36) & 55(40) \\ 
20 & 5 & $10^4$ & \textbf{25} & 61(27) & \cellcolor[HTML]{C0C0C0}56(27) & 55(29) & 60(31) & 60(31) \\ 
20 & 50 & $10^4$ & \textbf{100} & 50(44) & \cellcolor[HTML]{C0C0C0}51(47) & 51(45) & 56(29) & 60(31) \\ \noalign{\smallskip}\hline
\end{tabular}\footnotetext{Numbers in bold indicate the algorithm setting that is varied in each different scenario.}
\footnotetext{Grey-coloured column shows the best scaled version of $f(x)$}
\end{minipage}
\end{table}\
\\
Tables \ref{tab:5} and \ref{tab:7} show the best solution achieved, the mean best solution and standard deviation as well as the average number of function evaluation required to get a solution with a relative error $<5\%$ for \textit{E}6 and \textit{E}16, respectively. Tables \ref{tab:4} and \ref{tab:6} show the average number of function evaluations required to achieve the target values for \textit{E}4 and \textit{E}10 under different scaled versions. The results of the 50 trials are summarized in Fig. \ref{fig:5} and \ref{fig:6}. The computational efficiency is also measured by calculating the cumulative regret for each different scenario, see Fig. \ref{fig:7a} \& \ref{fig:7b}. The simple regret is here empirically calculated based on the trial which achieves the best solution.\\
\\
By increasing the number of iterations the mean best solution is closer to the true optimum, the standard deviation, as expected, becomes smaller and more trials succeed to get a solution within the acceptable level of error. But does the overall performance get better? Clearly, a small number of iterations can affect the precision of the results and the solution obtained can be far from the true global maximum (e.g. in \textit{E}4, \textit{E}6 and \textit{E}10). Considering the mean simple regret, the convergence to zero can be achieved but this happens only in a later stage as it requires more computational time. If the computational time is not a constraint, a higher number of iterations can be beneficial and add value to the entire optimization process but not always. For example, in \textit{E}4 and \textit{E}10, the more iterations performed the better the mean best solution is, whereas in \textit{E}6 and \textit{E}15, after a certain number of iterations there no substantial improvement and a lot of computational time is wasted.
\clearpage 
\begin{table}
\begin{minipage}{\textwidth}
\caption{Tunning Algorithm Settings: Sasena 2D (\textit{E}6). Brackets show the number of successful trials, out of the 50 performed, which achieved a solution with a relative error $<5\%$.}
\label{tab:5}
\begin{tabular}{|cccclrrc|}
\hline\noalign{\smallskip}
  T           & K           & Nsearch         & Ncand        & \multicolumn{1}{c}{\begin{tabular}[c]{@{}c@{}}Best\\ Solution\end{tabular}} & \multicolumn{1}{c}{\begin{tabular}[c]{@{}c@{}}Mean \\ Best \\ Solution \end{tabular}} & \multicolumn{1}{c}{SD} & \begin{tabular}[c]{@{}c@{}}Mean Function  \\ Evaluation \\ \textit{E} $< 5\%$\end{tabular} \\ \hline
20          & 5           & $10^4$           & 50           &  1.4564  & 1.4553  & 0.0012  & 57(50)      \\  
\textbf{10} & 5           & $10^4$           & 50           &  1.4523  & 1.0126  & 0.5270  & 37(9)       \\
\textbf{30} & 5           & $10^4$           & 50           &  1.4565  & 1.4566  & 0.0024  & 58(50)      \\ 
\textbf{50} & 5           & $10^4$           & 50           &  1.4565  & 1.4559  & 0.0031  & 58(50)      \\ 
20          & \textbf{2}  & $10^4$           & 50           &  1.4366  & 0.9441  & 0.7010  & 29(7)        \\
20          & \textbf{10} & $10^4$           & 50           &  1.4565  & 1.4561  & 0.0004  & 65(50)       \\ 
20          & \textbf{15} & $10^4$           & 50           &  1.4565  & 1.4561  & 0.0003  & 68(50)      \\ 
20          & 5           & \textbf{50}     & 50           &  1.4556  & 1.1801  & 0.8431  & 60(21)        \\ 
20          & 5           & \bm{$10^3$}   & 50           &  1.4560  & 1.3630  & 0.6103  & 55(47)       \\ 
20          & 5           & \bm{$10^5$} & 50           &  1.4565  & 1.4555  & 0.0012  & 62(50)        \\ 
20          & 5           & $10^4$           & \textbf{25}  &  1.4564  & 1.4551  & 0.0067  & 61(50)        \\ 
20          & 5           & $10^4$           & \textbf{100} &  1.4565  & 1.4540  & 0.0019  & 50(50)        \\  \noalign{\smallskip}\hline
\end{tabular}
\footnotetext{Numbers in bold indicate the algorithm setting that is varied in each different scenario.} 
\end{minipage}
\end{table} 

\begin{table}
\begin{minipage}{\textwidth}
\caption{Tunning Algorithm Settings: Scaling Rosenbrock 3D (\textit{E}10). Brackets show the number of successful trials, out of the 50 performed, which achieved a solution with a relative error $<5\%$.}
\label{tab:6}
\begin{tabular}{|ccccccccc|}
\hline\noalign{\smallskip}
 \multirow{5}{*}{T} & \multirow{5}{*}{K} & \multirow{5}{*}{Nsearch} & \multirow{5}{*}{Ncand} & \multicolumn{5}{c|}{Mean Function Evaluations (E \textless 5\%)} \\ \cline{5-9} 
\rule{0pt}{0.35cm} &  &  &  & \multicolumn{1}{c}{\multirow{4}{*}{\begin{tabular}[c]{@{}c@{}}Non-\\ Scaled \\ Version\end{tabular}}} & \multicolumn{4}{c|}{Scaled Versions} \\ \cline{6-9} 
 &  &  &  & \multicolumn{1}{c}{} & \multicolumn{2}{c}{Vertical} & \multicolumn{2}{c|}{Horizontal} \\ \cline{6-9} 
\rule{0pt}{0.5cm} &  &  &  & \multicolumn{1}{c}{} & $0.5f(x)$ & $2f(x)$ & $f(0.5x)$ & $f(2x)$ \\ \hline
30 & 5 & $10^4$           & 100  & 100(50) &\cellcolor[HTML]{C0C0C0}75(50) &112(42)  & 88(50) & 97(41)  \\ 
\textbf{20} & 5 & $10^4$  & 100  & 83(31)  &\cellcolor[HTML]{C0C0C0}74(45) & 84(29)  & 78(34) & 73(34)  \\ 
\textbf{40} & 5 & $10^4$  & 100  & 94(50)  &\cellcolor[HTML]{C0C0C0}78(50) & 121(50) & 87(50) & 99(50)  \\ 
\textbf{60} & 5 & $10^4$  & 100  & 99(50)  &\cellcolor[HTML]{C0C0C0}81(50) & 118(50) & 86(50) & 100(50)  \\ 
20 & \textbf{2} & $10^4$  & 100  & 47(5)   &\cellcolor[HTML]{C0C0C0}53(11) & 46(2)   & 54(5)  & 43(4)  \\ 
20 & \textbf{10} & $10^4$ & 100  & 114(50) &\cellcolor[HTML]{C0C0C0}82(50) & 138(50) & 102(50)& 118(50)  \\ 
20 & \textbf{15} & $10^4$ & 100  & 115(50) &\cellcolor[HTML]{C0C0C0}97(50) & 159(50) & 104(50)& 137(50)  \\ 
20 & 5 & \bm{$100$}  & 100  & 94(12)  &\cellcolor[HTML]{C0C0C0}81(29) & 98(4)   & 79(15) & 82(18)  \\
20 & 5 & \bm{$10^3$}   & 100  & 97(41)  &\cellcolor[HTML]{C0C0C0}87(49) & 112(26) & 98(45) & 92(41)  \\ 
20 & 5 & \bm{$10^5$} & 100  & 96(50)  &\cellcolor[HTML]{C0C0C0}79(50) & 114(48) & 85(50) & 101(49)  \\ 
20 & 5 & $10^4$ & \textbf{50}    & 95(50)  &\cellcolor[HTML]{C0C0C0}83(50) & 114(35) & 90(50) & 99(48)  \\ 
20 & 50 & $10^4$ & \textbf{150}  & 95(50)  &\cellcolor[HTML]{C0C0C0}79(50) & 113(42) & 79(50) & 105(47)  \\ \noalign{\smallskip}\hline
\end{tabular}\footnotetext{Numbers in bold indicate the algorithm setting that is varied in each different scenario.}
\footnotetext{Grey-coloured column shows the best scaled version of $f(x)$}
\end{minipage}
\end{table}\
\\
The amount of exploration needed is not known beforehand but it can be controlled by the size of the batch. Using a small batch size, the uncertain regions are not fully explored and the mean best solution is far from the optimum (e.g. \textit{E}4, \textit{E}10). In this case, there is a high possibility to fail to achieve the target value (e.g. when $K=2$, Tables 4-7), or the regret decays after performing a lot of function evaluations and often fails to converge into zero (Fig. \ref{fig:7a} \& \ref{fig:7b} ). On the other hand, a large batch size helps the algorithm in the search process in the very first few iterations, where the knowledge about the unknown function is limited, and always ensures a good solution. As more input points are chosen in each iteration, the regret decays straight away indicating that the uncertain region can be found in few function evaluations (Fig. \ref{fig:7a} \& \ref{fig:7b}).\\ 
\\
But having a larger batch size does not always guarantee a massive improvement of the overall performance of the optimization. For example, in \textit{E}4 and \textit{E}10, a larger batch size gets a solution even closer to the true optimum but not as close as if more iterations would be performed, in \textit{E}6, a batch of 10 or 15 input points does not make substantial difference whereas in \textit{E}15, a batch of 5, 10 or 15 input points yields almost the same mean best solution. Once the uncertain region is found, the progression can be slow and the convergence to zero can only be achieved after a lot of function evaluations (Fig. \ref{fig:7a} \& \ref{fig:7b}). A possible reason might be that the explored region is tiny and the new candidate points drawn are almost identically which makes it much harder to see an improvement. What it is also worth mentioning is that by increasing the batch size, the number of function evaluations required to achieve the target value is also increased, and compared with the scenarios where the number of iterations is increased, that number is much bigger. This indicates that exploring the search space more than what is needed is not always computationally efficient despite the better solution that can be achieved.\\
\\
Although optim-MICE is computationally more expensive than the other two alternative schemes, it is more effective. The quality of the points added in the design at each time step is better and as a result, the true optimum is achieved in less function evaluations. Due to the limited computational resources, MICE is only examined on a certain number of candidate points. If the number of potential design points included in the search space is small and the same as the number of candidate points (e.g. $Nsearch=50$ and $Ncand=50$, Table \ref{tab:4}) which are used for the exploration part, the uncertain regions are not fully explored, the solution is often far from the true optimum and the regret converges to zero after a lot of function evaluations. There is also higher possibility to fail to achieve a solution with a relative error $<5\%$ because, considering the results obtained in the four experiments (Tables 4-7), only half of the 50 trials performed are successful. In contrast, as it is expected, having a high $Nsearch$, the target value is found in less computational time and the decay of regret is done faster.\\
\\
The number of candidate points examined by MICE is a vital setting to the overall performance of the algorithm. Despite the fact that in the ALM-based algorithm a much larger number of candidate points are examined at each iteration, the results in the computational experiments show that the MICE-based algorithm still performs better even with a smaller candidate set. Considering the results obtained in the four test cases, a larger set of candidate points gives a solution even closer to the true optimum with the lowest possible number of function evaluations. The proposed algorithm performs also well with a small candidate set, as the mean best solution is still within an acceptable distance from the true optimum, however, with a larger set, the average number of function evaluation required to get a solution with a relative error $<5\%$ is much lower (Tables 4-7). How fast the regret decays depends on the complexity of the function, but having a large number of candidate points allows the convergence of regret to zero in fewer function evaluations (Fig. \ref{fig:7a} \& \ref{fig:7b}).

\begin{table}
\centering
\begin{minipage}{\textwidth}
\caption{Tunning Algorithm Settings: Hartman 6D (\textit{E}15). Brackets show the number of successful trials, out of the 50 performed, which achieved a solution with a relative error $<5\%$.}
\label{tab:7}
\begin{tabular}{|cccclrrc|}
\hline\noalign{\smallskip}
  T           & K           & Nsearch         & Ncand        & \multicolumn{1}{c}{\begin{tabular}[c]{@{}c@{}}Best\\ Solution\end{tabular}} & \multicolumn{1}{c}{\begin{tabular}[c]{@{}c@{}}Mean \\ Best \\ Solution \end{tabular}} & \multicolumn{1}{c}{SD} & \begin{tabular}[c]{@{}c@{}}Mean Function  \\ Evaluation \\ \textit{E} $< 5\%$\end{tabular} \\ \hline
60          & 5           & $10^4$           & 250          &  3.286   & 3.228  & 0.023  & 106(49)      \\
\textbf{50} & 5           & $10^4$           & 250          &  3.260   & 3.191  & 0.031  & 114(49)       \\ 
\textbf{70} & 5           & $10^4$           & 250          &  3.296   & 3.229  & 0.019  & 117(50)      \\
\textbf{90} & 5           & $10^4$           & 250          &  3.296   & 3.235  & 0.014  & 113(50)      \\  
60          & \textbf{2}  & $10^4$           & 250          &  3.250   & 3.188  & 0.058  & 72(41)        \\ 
60          & \textbf{10} & $10^4$           & 250          &  3.292   & 3.231  & 0.021  & 143(50)       \\ 
60          & \textbf{15} & $10^4$           & 250          &  3.292   & 3.230  & 0.020  & 170(50)      \\ 
60          & 5           & \textbf{250}    & 250          &  3.241   & 3.158  & 0.036  & 156(24)        \\ 
60          & 5           & \bm{$10^3$}   & 250          &  3.259   & 3.189  & 0.031  & 130(40)       \\ 
60          & 5           & \bm{$10^5$} & 250          &  3.294   & 3.258  & 0.011  & 86(50)        \\ 
60          & 5           & $10^4$           & \textbf{200} &  3.263   & 3.225  & 0.024  & 110(48)        \\
60          & 5           & $10^4$           & \textbf{300} &  3.285   & 3.230  & 0.022  & 101(50)        \\ \noalign{\smallskip}\hline
\end{tabular}
\footnotetext{Number in bold indicate the algorithm setting that is varied in each different scenario.} 
\end{minipage}
\end{table}\
\\
Compared with the non-scaled version, scaling \textit{E}4 either vertically or horizontally does not make a substantial difference in the solution achieved. Overall, the best solutions achieved from all the 50 trials in each scaled-version, shown in box-plots in Fig. \ref{fig:5}, are very close to the solutions achieved in the non-scaled version. In the scaled-versions, the mean best solution obtained in each scenario is almost the same with the vertical scaling by 0.5 ($y=0.5f(x)$) and 2 ($y=2f(x)$) being slightly better than the others. As in the non-scaled version, in the scaled ones, by increasing the number of iterations and the batch size and choosing the design points from a larger candidate set, the computational efficiency is improved. What is really achieved by scaling vertically the function in \textit{E}4, especially by 0.5, is that the mean function evaluations required to get a solution with relative error $<5\%$ is lower than in the other examined versions (Table \ref{tab:4}). Therefore, the true optimum can be achieved in less computational time.\\
\\
The advantage of scaling the objective function is better shown in \textit{E}10. Fig. \ref{fig:6}, which summarizes the best solutions achieved from all the all trials, shows that scaling the function by a small factor such as 0.5, either vertically ($y=0.5f(x)$) or horizontal ($y=f(0.5x)$), the overall performance of the algorithm is more effective. In both versions, the variation between the best solution achieved in each trial is smaller and the number of successful trials is bigger compared to the other versions and this gives even more certainty about the algorithm's performance. However, computational efficiency is clearly achieved when the function is vertically scaled by 0.5. In all the different scenarios, the mean best solution is closer to the true optimum and the mean function evaluations required to obtain the target value is lower than in any other version.     

\section{Conclusion}
Overall, incorporating the MICE criterion in the GP framework seems to be more computationally beneficial compared to the other two approaches. The optim-MICE method increases the confidence in getting a solution close to the true optimum, in less function evaluations. Regardless of the dimensionality and complexity of the computer model, with optim-MICE the search process for identifying the uncertain region is more efficient and the regret decays faster compared to the alternatives. The total computational time needed to find the true optimum and achieve convergence of the regret is less than the others approaches.\\
\\
The performance of the proposed optimization scheme is affected by the algorithm settings. A complex and a high-dimensional computer model clearly needs more iterations and a bigger batch size than a simple and a low-dimensional function. To get the most of the proposed optimization scheme, a balance between the number of iterations and the batch size is needed according to the complexity that a function might have. If one performs a lot of iterations without exploring enough, the information gain about the unknown function at each time step is limited as the number of input points added in the design is small and the objective function is evaluated only a few times. Furthermore, the advantage of MICE is not fully utilized and a certain computational cost is added, without necessarily needed, such as the re-estimation of the hyper-parameters of the surrogate model. \\
\\
When making us of a large batch size, the uncertain region is definitely explored more and the chance to find to true optimum in fewer function evaluations is higher. But, choosing a larger batch size than what is needed, an amount of computational time is wasted as the algorithm is forced to stay in a region which might not be of interest anymore - and has already discovered it from the first exploration steps - or in a region which has already been explored and any additional information will not add value. In both cases, it is  unavoidable that a number of function evaluations are performed without obtaining any progress. To keep the number of function evaluations as low as possible when a large batch size is chosen, it is worth scaling the objective function vertically by a small factor (e.g. 0.5). In terms of $Nsearch$ and $Ncand$, things are more straightforward. Regardless of dimensionality and complexity of the function, having a large number of input points spread around the search space and choosing to examine a big set of candidate points with the MICE criterion could lead us to more accurate results without wasting more computational resources. \ 
\\

\begin{landscape}
\begin{subfigures}
\begin{figure}[ht]
\centering
\minipage{0.7\textwidth}
  \includegraphics[width=1\linewidth]{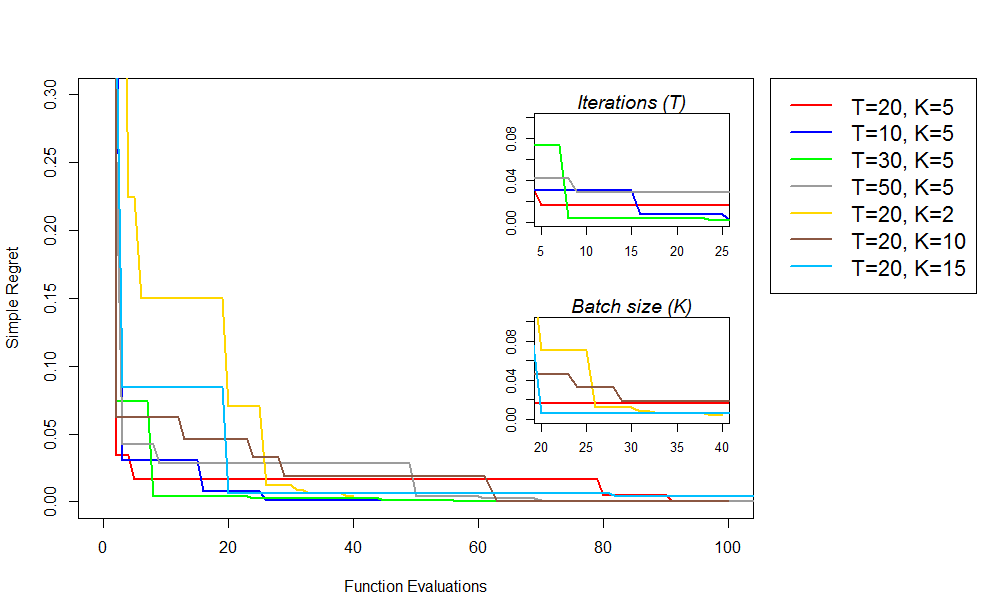}
\endminipage
\minipage{0.7\textwidth}
  \includegraphics[width=1\linewidth]{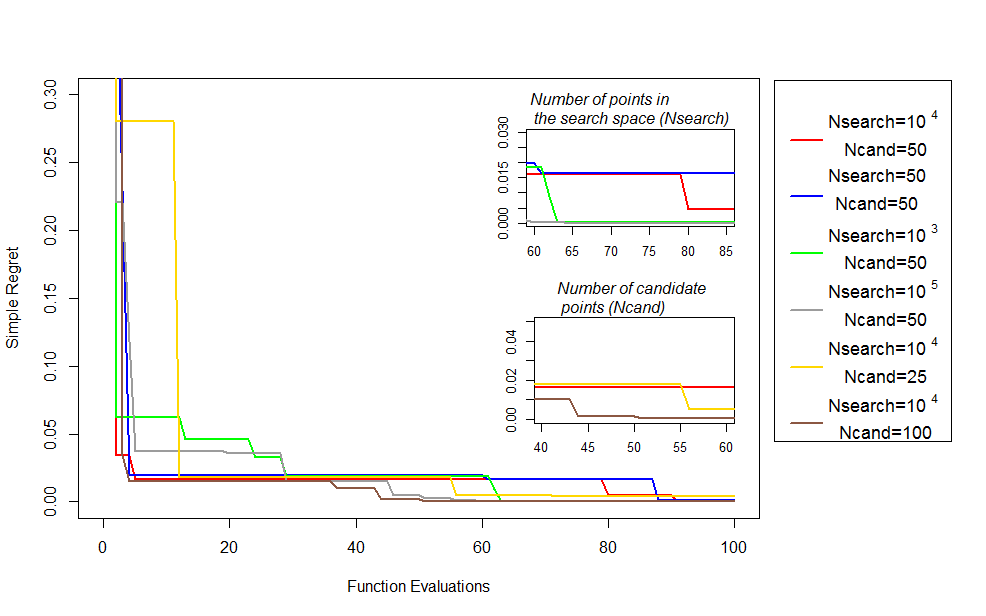}
\endminipage\\
\centering\footnotesize{Hosaki 2D (\textit{E}4)}\\
\minipage{0.7\textwidth}
  \includegraphics[width=1\linewidth]{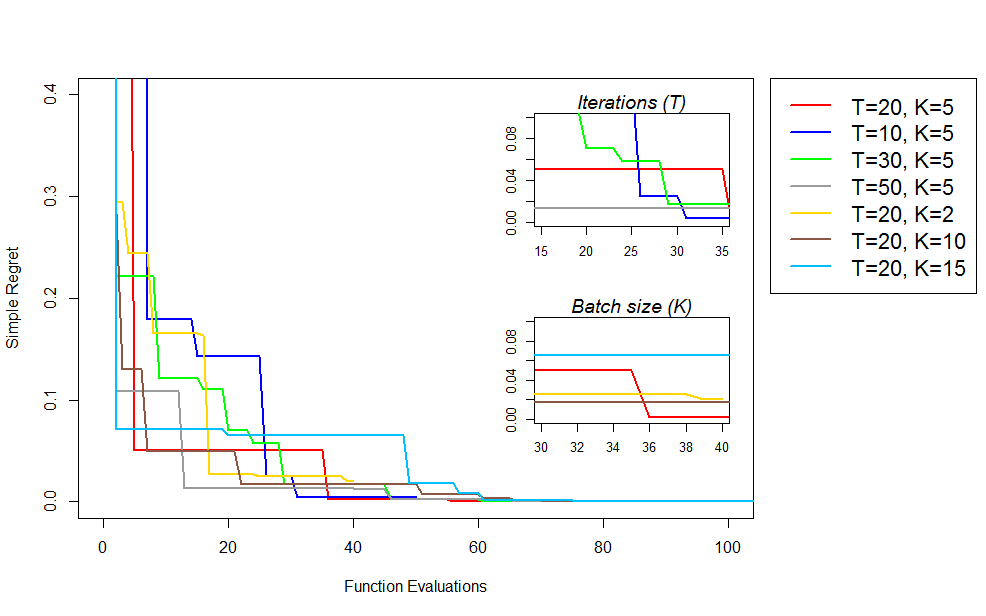}
\endminipage
\minipage{0.7\textwidth}
  \includegraphics[width=1\linewidth]{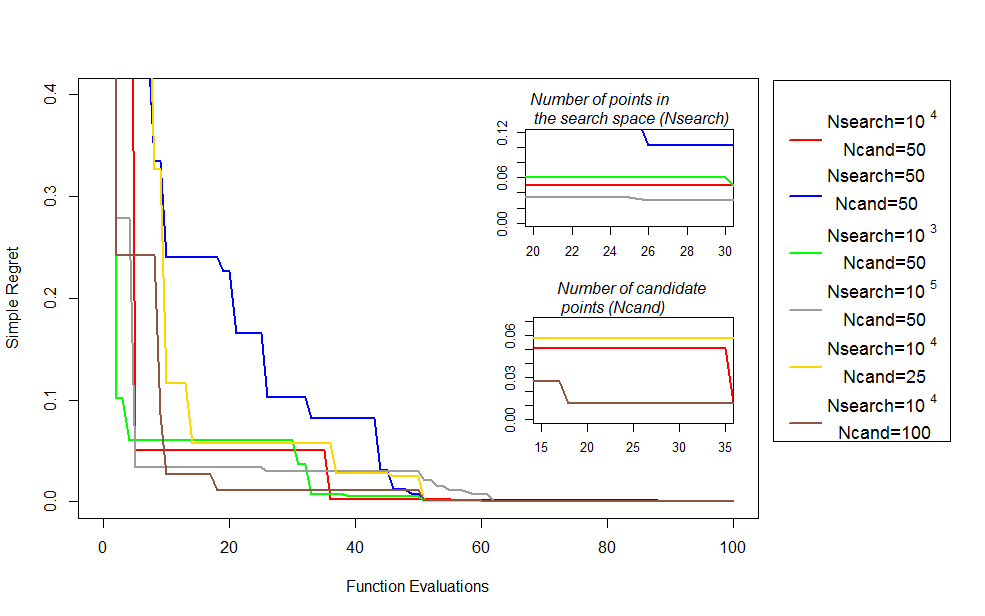}
\endminipage\\
\centering\footnotesize{Sasena 2D (\textit{E}6)}
\captionof{figure}{Mean simple regret combinations of the algorithm settings. \textit{Left}: Number of iterations (\textit{T}) and Batch size (\textit{K}). \textit{Right}: Number of points in the search space (\textit{Nsearch}) and Number of candidate points (\textit{Ncand}).}
\label{fig:7a} 
\end{figure}

\begin{figure}[ht]
\ContinuedFloat
\centering
\minipage{0.7\textwidth}
  \includegraphics[width=1\linewidth]{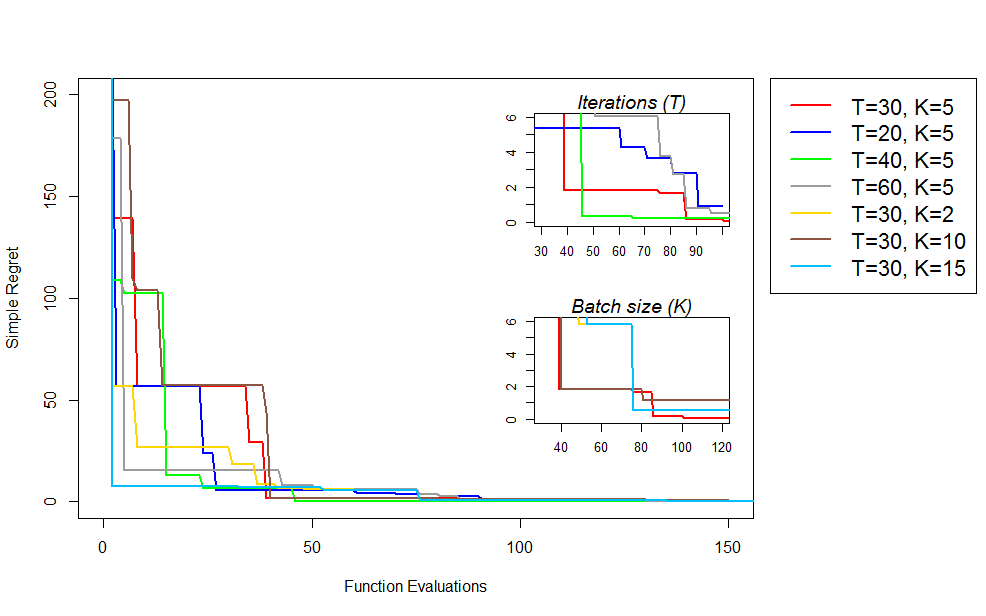}
\endminipage
\minipage{0.7\textwidth}
  \includegraphics[width=1\linewidth]{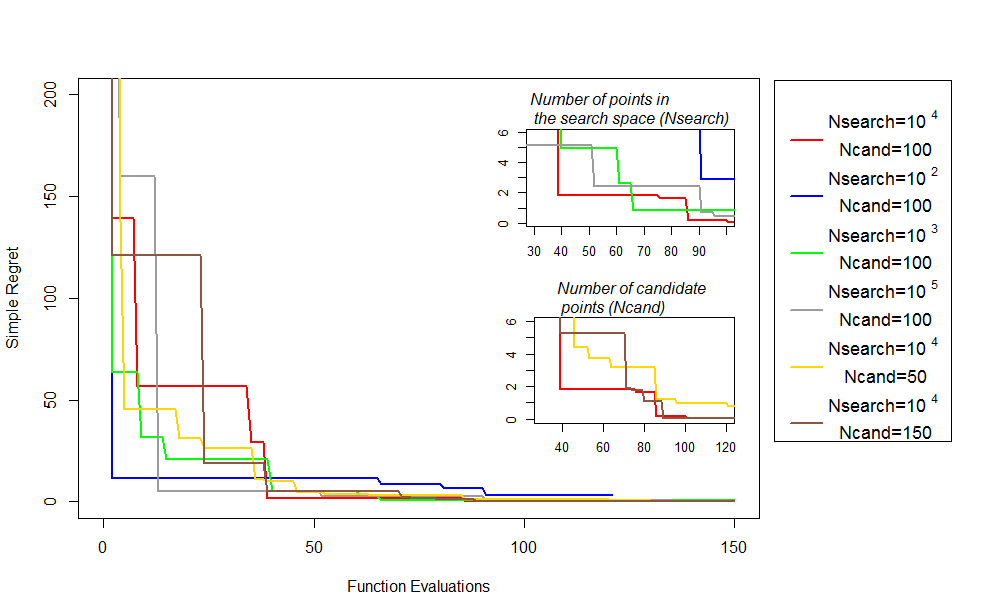}
\endminipage\\
\centering\footnotesize{Rosenbrock 3D (\textit{E}10)}\\
\minipage{0.7\textwidth}
  \includegraphics[width=1\linewidth]{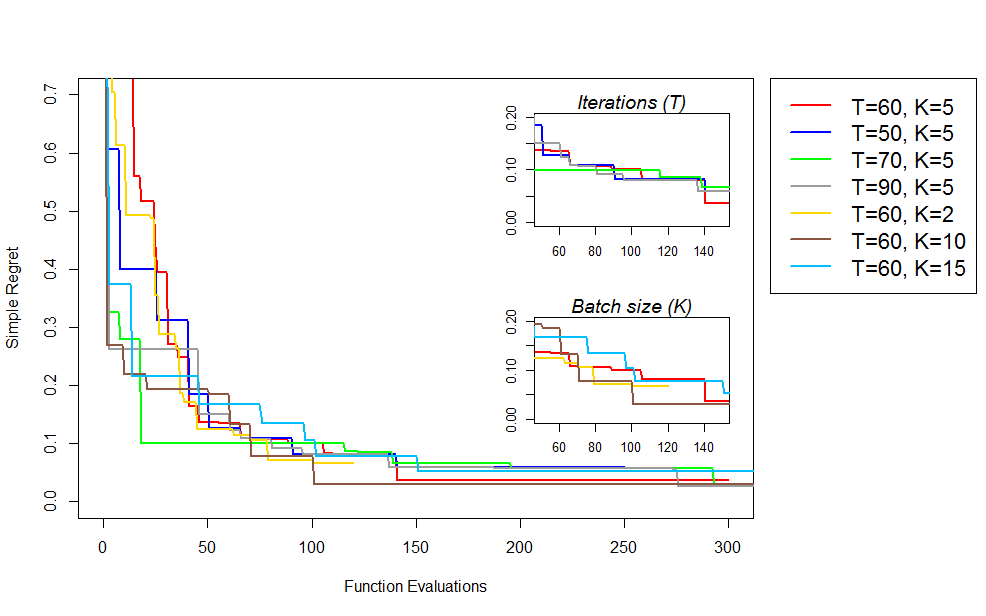}
\endminipage
\minipage{0.7\textwidth}
  \includegraphics[width=1\linewidth]{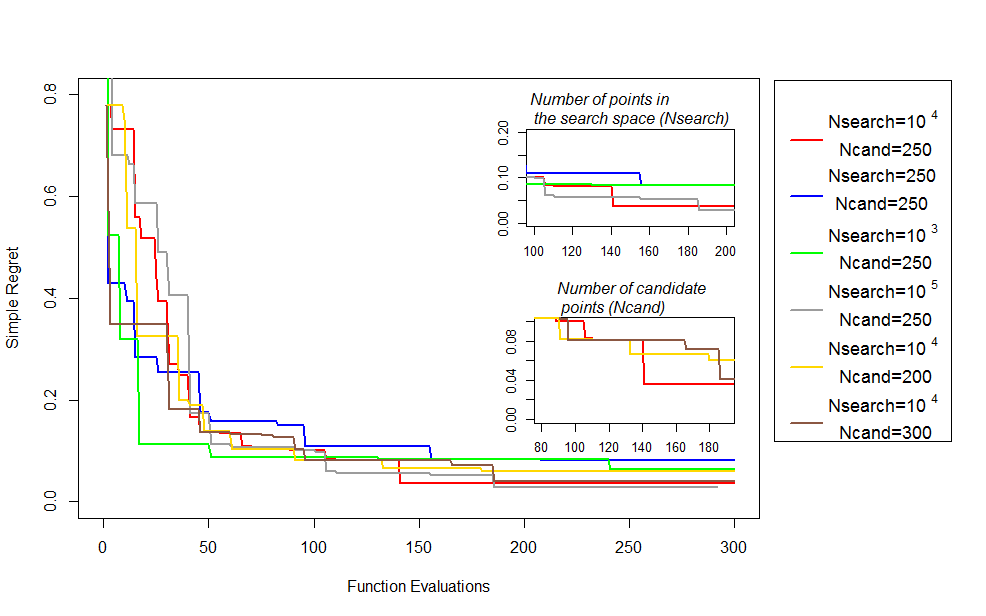}
\endminipage\\
\centering\footnotesize{Hartmann-6 6D (\textit{E}15)}
\captionof{figure}{Mean simple regret combinations of the algorithm settings. \textit{Left}: Number of iterations (\textit{T}) and Batch size (\textit{K}). \textit{Right}: Number of points in the search space (\textit{Nsearch}) and Number of candidate points (\textit{Ncand}).}
\label{fig:7b} 
\end{figure}
\end{subfigures}
\end{landscape}

\clearpage
\bibliographystyle{abbrv}      
\bibliography{main}  
\end{document}